\documentclass[]{aa}
\usepackage{amsmath}
\usepackage{graphicx}
\usepackage{tabularx}
\usepackage{txfonts}
\usepackage{aas_macros}
\usepackage{calc}
\usepackage{aalongtable}
\usepackage{natbib}

\begin{document}
   \title{The VIMOS VLT Deep Survey: the faint type-1 AGN sample
\thanks{Based on data obtained with the European Southern Observatory
  Very Large Telescope, Paranal, Chile, program 070.A-9007(A),
  272.A-5047, 076.A-0808 and on data obtained at the Canada-France-Hawaii
  Telescope, operated by the CNRS of France, CNRC in Canada and the
  University of Hawaii.}} 

\author{
I. Gavignaud \inst{1,2}
\and A. Bongiorno \inst{3}
\and S. Paltani \inst{4,5}
\and G. Mathez \inst{2}
\and G. Zamorani \inst{6} 
\and P. M\o ller\inst{7}
\and J.P. Picat \inst{2}
\and V. Le Brun \inst{8}
\and B. Marano \inst{3}  
\and O. Le F\`evre \inst{8}
\and D. Bottini \inst{9}
\and B. Garilli \inst{9}
\and D. Maccagni \inst{9}
\and R. Scaramella \inst{10,11}
\and M. Scodeggio \inst{9}
\and L. Tresse \inst{8}
\and G. Vettolani \inst{10}
\and A. Zanichelli \inst{10}
\and C. Adami \inst{8}
\and M. Arnaboldi \inst{12}
\and S. Arnouts \inst{8}
\and S. Bardelli  \inst{6}
\and M. Bolzonella  \inst{6} 
\and A. Cappi    \inst{6}
\and S. Charlot \inst{13,14}
\and P. Ciliegi    \inst{6}  
\and T. Contini \inst{2}
\and S. Foucaud \inst{9}
\and P. Franzetti \inst{9}
\and L. Guzzo \inst{15}
\and O. Ilbert \inst{3}
\and A. Iovino \inst{15}
\and H.J. McCracken \inst{14,16}
\and C. Marinoni \inst{8}
\and A. Mazure \inst{8}
\and B. Meneux \inst{9,15}
\and R. Merighi   \inst{6} 
\and R. Pell\`o \inst{2}
\and A. Pollo \inst{8}
\and L. Pozzetti    \inst{6} 
\and M. Radovich \inst{12}
\and E. Zucca    \inst{6}
\and M. Bondi \inst{10}
\and G. Busarello \inst{12}
\and O. Cucciati \inst{15,17}
\and S. de la Torre \inst{8}
\and L. Gregorini \inst{10}
\and F. Lamareille \inst{2}
\and Y. Mellier \inst{14,16}
\and P. Merluzzi \inst{12}
\and V. Ripepi \inst{12}
\and D. Rizzo \inst{2}
\and D. Vergani \inst{9}
}

\institute{
%1) 
Astrophysikalisches Institut Potsdam, An der Sternwarte 16, D-14482 Potsdam,
Germany
\and
%2) 
Laboratoire d'Astrophysique de l'Observatoire Midi-Pyr\'en\'ees (UMR 
5572) -
14, avenue E. Belin, F31400 Toulouse, France
\and
%3) 
Universit\`a di Bologna, Dipartimento di Astronomia - Via Ranzani,1,
I-40127, Bologna, Italy
\and
%4) 
Integral Science Data Centre, ch. d'\'Ecogia 16, CH-1290 Versoix
\and
%5) 
Geneva Observatory, ch. des Maillettes 51, CH-1290 Sauverny
\and
%6) 
INAF-Osservatorio Astronomico di Bologna - Via Ranzani,1, I-40127, Bologna, Italy
\and
%7)
European Southern Observatory, Karl-Schwarzschild-Strasse 2, D-85748
Garching bei München, Germany
\and
%8)
Laboratoire d'Astrophysique de Marseille, UMR 6110 CNRS-Universit\'e de
Provence,  BP8, 13376 Marseille Cedex 12, France
\and
%9)
IASF-INAF - via Bassini 15, I-20133, Milano, Italy
\and
%10)
IRA-INAF - Via Gobetti,101, I-40129, Bologna, Italy
\and
%11) 
INAF-Osservatorio Astronomico di Roma - Via di Frascati 33,
I-00040, Monte Porzio Catone,
Italy
\and
%12) 
INAF-Osservatorio Astronomico di Capodimonte - Via Moiariello 16, I-80131, Napoli,
Italy
\and
%13) 
Max Planck Institut fur Astrophysik, 85741, Garching, Germany
\and
%14-n) 
Institut d'Astrophysique de Paris, UMR 7095, 98 bis Bvd Arago, 75014
Paris, France
\and
%15) 
INAF-Osservatorio Astronomico di Brera - Via Brera 28, Milan,
Italy
\and
%16-o) 
Observatoire de Paris, LERMA, 61 Avenue de l'Observatoire, 75014 Paris, 
France
\and
%17) 
Universit\'a di Milano-Bicocca, Dipartimento di Fisica - 
Piazza delle Scienze, 3, I-20126 Milano, Italy
}

\offprints{I. Gavignaud, \email{igavignaud@aip.de}}

      \date{Received ..., ....; accepted ..., ....}

      \abstract{
        We present the type-1 active galactic nuclei (AGN) sample
        extracted from the VIMOS VLT Deep Survey first observations
        of 21000 spectra in 1.75 deg$^2$. 
        This sample, which is purely magnitude limited, free of morphological
        or color selection biases, contains 130 broad line AGN (BLAGN) spectra
        with redshift up to 5. 
        Our data are divided into a wide ($I_{AB} \le 22.5$) and a
        deep ($I_{AB} \le 24$) subsample containing 56 and 74 objects
        respectively.
        Because of its depth and selection criteria, this sample is
        uniquely suited to study the population of faint type-1 AGN.
        Our measured surface density ($\sim 472 \pm 48$ BLAGN per square
        degree with $I_{AB} \le 24$) is significantly higher than that of any 
        other optically selected sample of BLAGN with spectroscopic
        confirmation.   
        By applying a morphological and color analysis to our AGN
        sample we find that:
        (1) $\sim 23 \%$ of the AGN brighter than $I_{AB}=22.5$ are
        classified as extended; this percentage increases to $\sim$ 
        42 \% for those with $z < 1.6$;
        (2) a non-negligible fraction of our BLAGN are lying close to
        the color space area occupied by stars in $u^*-g'$ versus $g'-r'$
        color-color diagram.
        This leads us to the conclusion that classical optical ultraviolet
        preselection technique, if employed at such deep magnitudes
        ($I_{AB}=22.5$) in conjuction with a preselection of point-like
        sources, can miss miss up to $\sim 35\%$ of the AGN population.  
        Finally, we present a composite spectrum of our sample of
        objects.
        While the continuum shape is very similar to that of
        the SDSS composite at short wavelengths, it is much redder than
        it at $\lambda \ge 3000$ \AA . We interpret this as due to
        significant contamination from emission of the host galaxies,
        as expected from the faint absolute magnitudes sampled  by our
        survey.
      }
     \authorrunning
     \titlerunning
     \maketitle

%
%________________________________________________________________

\section{Introduction}

Recent surveys such as 2QZ and SDSS produce quasi-stellar object (QSO) spectra
by the thousands \citep{Croom2004,Richards2002}, 
demonstrating  the high efficiency of optical color selection
techniques used since the pioneering work by \citet{Sandage1965}.
Spectroscopic targets are preselected based
on their location in multidimensional color space (far from main sequence
stars) and their morphology (point-like appearance, to avoid galaxies). 
The drawback  of such prerequisites is that some subsets of the underlying
QSO population may be under-represented in these surveys, thereby
possibly biasing our current understanding of this population of
objects. 

The preselection of non-resolved objects  prevents the selection of
faint active nuclei in relatively bright galaxies and introduces incompleteness
at low redshift.
Moreover, the selection of unresolved objects is highly dependent on the
quality of the imaging data.
The ultraviolet (UV) excess selection is efficient only up to $z\sim2.3$, and
the evolving track of standard QSO crosses over main sequence stars near $z
\sim 2.5 \pm 0.5$ in most optical broad band color-color planes.   
While it is difficult to identify AGN candidates in this redshift
range, it is also at this epoch that a maximum in the 
QSO space density seems to be observed \citep{Wall2005}, although there is now clear
indication, at least for the X-ray selected AGN, that this maximum is
dependent on the intrinsic luminosity \citep{Hasinger2005}. 
It is therefore essential to obtain unbiased samples at these high
redshifts to trace and understand the accretion history onto
supermassive black holes, which is believed to be the origin of the
AGN phenomena. 

Other criteria have been used to select AGN samples such as
spectral energy distribution of objects in 
slitless spectroscopic plates \citep{Hewett1995,Wisotzki2000},
variability and low proper motion \citep{Brunzendorf2002} or radio emission
\citep{Jackson2002,Ivezic2002}, but they are always biased toward some subsets
of the population.
X-ray surveys are an efficient alternative to traditional optically
selected samples and, indeed, they have allowed
to derive, with good statistics, luminosity function and evolution of
various classes of X-ray emitting AGN (e.g. absorbed and unabsorbed
AGN in the X-ray band, soft and hard X-ray selected, 
see \citealp{Miyaji2000,Ueda2003,Hasinger2005}). 
However, the comparison of the results obtained for X-ray and
optically selected AGN is not straightforward both because of the different
composition of the samples (X-ray samples contain large fraction of
obscured AGN, without broad emission lines in their optical spectra)
and of the somewhat different definition of AGN which is usually adopted
in the X-ray surveys (e.g. any source with an X-ray luminosity and/or 
an X-ray to optical luminosity ratio above given thresholds).

The identification of faint type I AGN in a complete magnitude limited optical
sample has been attempted in the Canada France Redshift Survey (CFRS) with 6 QSO
identified among 943 spectra down to a limiting magnitude of $I_{AB} = 22.5$
\citep{Schade1996}. 

To probe fainter magnitude the COMBO-17 project identifies AGN from their
spectral energy distribution (SED)  based on photometry in
a set of 17 filters. They selected 192 objects over 0.78 square degree down to
a magnitude $R = 24$ with $z >1.2$ \citep{Wolf2003}. 
Although the technique developed by this team leads to a good
redshift accuracy ($\Delta z\sim0.03$), it can be affected by some
'catastrophic errors' in region of the color space where various
identifications are possible or in the case of objects with SED deviating from
the reference set of templates. 

In the present paper we present a spectroscopic catalog of broad emission line
objects obtained from the purely flux limited spectroscopic sample of the
VIMOS VLT Deep Survey (VVDS).  
This sample, free of any color or morphological biases, can be considered as a
minimally biased sample to explore the population of optically non-obscured
QSO at the faint end of the luminosity function. 
Our completeness is essentially set by the S/N ratio at which a broad
emission line can be reliably identified in our spectra.  

The data are divided into a wide ($I_{AB} \le 22.5$) and a
deep ($I_{AB} \le 24$) subsample containing 56 and 74 objects
respectively.
The limiting magnitude of our deep sample is well beyond the limit of
other optical spectroscopic surveys of AGN.

In the following section we describe briefly the VVDS imaging and
spectroscopic surveys. In Section \ref{Catalog} and \ref{SF} we describe the
process of AGN identification in the VVDS and the corresponding selection
function.  
In Sections \ref{Counts}, \ref{Zdistrib} and \ref{Morphcol} we present the
surface density of 
BLAGN, the redshift distribution and the morphological and color
properties of our sample. The composite spectrum of our
sample of objects is presented and discussed in section \ref{sec:composite}.

%====================================================================
%   Observations
%
%====================================================================

\section{Observations}

The VVDS \footnote{http://www.oamp.fr/virmos/vvds.htm} is a faint imaging
and $I_{AB}$ magnitude limited spectroscopic survey. The spectroscopic
VVDS survey consists of the VVDS-deep,
targeting objects in the range $I_{AB}=17.5-24.0$, and the VVDS-wide, 
targeting objects in the range $I_{AB}=17.5-22.5$.

The purely magnitude-limited selection of spectroscopic targets
was carried out from a catalog based on deep photometry. This
catalog is complete down to $I_{AB}=25$ in all fields. In addition,
deep multi-band photometry was obtained, providing comprehensive
multi-wavelength information for the objects.
$B,V,R$ and $I$ photometry was
obtained on the full fields of view and $U$, $K$ and $J$ in smaller areas.
For a detailed description of the photometric catalog we
refer the reader to \cite{LeFevre2004a}.
A comprehensive description of the data reduction and final data
quality of the deep imaging survey can be found in \cite{McCracken2003}.

Spectroscopic observations were performed with the multi-object
spectrograph VIMOS installed on Melipal \citep[see][]{LeFevre2003}, the third 8-meter
telescope at Paranal Observatory.
In this paper we present results based on the first
spectroscopic runs only (Epoch One), obtained during October and
November 2002. These observations and the reduction procedure are
presented in detail in \citet{LeFevre2005} for the deep survey and in
\citet{Garilli2006} for the wide sample.

\subsection{The VVDS multi-wavelength imaging survey}
%_________________________________________________ Imaging deep
\begin{table}
      \caption[]{Summary of some of the spectroscopic data
      available on VVDS first epoch fields.}
         \label{Imaging}
\begin{center}
\begin{tabular}{llrr}
   \hline \hline
   \noalign{\smallskip}
   Field       &  Mode   & TSR$^{*}$   & $\mathcal{A}^{**}$ \\
               &         &             & [deg$^{2}$]\\
   \noalign{\smallskip}
   \hline \hline
   \noalign{\smallskip}
   VVDS-0226-04& deep & 25\% & 0.48\\
   \noalign{\smallskip}
   \hline
   VVDS-1003+01& wide & 26\% & 0.33\\
   \noalign{\smallskip}
   \hline
   VVDS-2217+00& wide & 22\% & 0.81\\
   \noalign{\smallskip}
   \hline
   CDFS        & deep  & 23\% & 0.13\\
   \noalign{\smallskip}
   \hline \hline
\end{tabular}
\end{center}
\begin{list}{}{}
\item[$^{*}$] Target Sampling Rate: Fraction of objects in the
  photometric catalog inside our targeted area $\mathcal{A}$ which
  have been spectroscopically observed.  
\item[$^{**}$] Geometrical area in deg$^{2}$ of the spectroscopic
  first epoch data which are used in this work.
\end{list}
\label{tab:specdata}
\end{table}
%--------------------------------------------------

The wide and deep fields were imaged in four bands, $B, V, R$ and $I$.
The wide photometric survey covers 12 deg$^2$ in three high
galactic latitude fields, each 2x2 deg$^2$: VVDS-1003+01,
VVDS-1400+05, and VVDS-2217+00. The deep photometric survey
covers a 1.3 x 1 deg area, VVDS-0226-04. Part of the
deep field has been observed in additional
bands: $U$ \citep{Radovich2004}, $J$ and $K$ \citep{Iovino2005},
$u^*$,$g'$,$r'$,$i'$,$z'$ from the Canada-France-Hawaii Legacy Survey
(CFHTLS) \footnote{www.cfht.hawaii.edu/Science/CFHLS} as 
well as in radio (1.4 GHz) on the VLA \citep{Bondi2003}, X-ray with
XMM \citep{Pierre2004}, and UV with Galex
\citep{Arnouts2005,Schiminovich2005}.  

In order to avoid surface brightness selection biases down to the
faintest required spectroscopic limit of $I_{AB}=24$,
the imaging survey was designed to be complete down to
$I_{AB}=25$ \citep[see][]{McCracken2003}. 
The B, V, R and I observations were performed from
November 1999 to October 2000 at the 3.6 m Canada-France-Hawaii
Telescope, with the CFH12K wide field mosaic camera.

In addition to the VVDS fields, a deep spectroscopic
survey has also been performed in 0.13 deg$^2$ of the Chandra
Deep Field South (CDFS) field and is included in this work. 
This field is based on the EIS $I$ band catalog described in
\citet{Arnouts2001}. 
The spectroscopic data available in the CDFS are public and are
described in \cite{LeFevre2004a}. 

All magnitudes given in this paper are corrected for galactic
extinction using the dust map provided by \cite{Schlegel1998}.

\subsection{\label{selection} Target selection}

To get a fully unbiased spectroscopic sample one would
pick objects at random from the input catalog. Such a procedure would
ensure randomness, but would provide spectra at a less than optimal rate for
a given observing time.
For this reason a special target selection algorithm SPOC
was developed, with the aim of optimizing the use of spectroscopic
follow-up time given the constraints of the VIMOS optical and
mechanical layout \citep{Bottini2005}. In this optimization
process, however, two biases are introduced.

First, objects with a small projected size along the slit
are slightly favored for spectroscopic selection, resulting
in a bias against more extended objects. This bias has
to be taken into account in our analysis (as described in section
\ref{TSR}).

Second, the distribution of selected targets is not isotropic as SPOC prevents
the observation of targets closer than 2\arcmin\, perpendicular to the slit
direction to avoid surperpositions of spectra. 
However this bias is important and must be taken into account only for studies
related to correlation function and large-scale
structure \citep[see][]{Pollo2005} and can be ignored here.

\subsection{Spectroscopy}

The area covered by the Epoch One VVDS spectroscopy
is $\sim$0.6 deg$^2$ in the deep fields and $\sim$1.1 deg$^2$ in the
wide fields (see column 4 in Table \ref{tab:specdata}) and the
corresponding numbers of spectra are 11 564 and 9440 respectively.
The target sampling rate, i.e. the ratio of selected objects
for spectroscopy to the total number of objects down to the
respective spectroscopic flux limits, is $\sim$ 25\%
in both the wide and the deep surveys.

We used VIMOS and the LRRED grism which covers the wavelength range 5500 --
9500\AA\ with a 7\AA/pixel dispersion. The slit width was fixed to
1\arcsec, providing a sampling of 5 pixels per slit and a
spectral resolution of $R \sim 230$ %= 227, or better at 7500 \AA ,
corresponding to a spectral resolution element of $\sim 33$ \AA.
The total integration time is 4.5 hours per mask in the deep fields
and 50 minutes in the wide fields.
Data reduction was performed homogeneously 
using the VIPGI software which was developed specifically for the VVDS survey
\cite{Scodeggio2005}. 

The measured median signal-to-noise ratio (S/N) per resolution
 element at our limiting magnitudes is 4 in both the deep and wide fields. 
This is sufficient to detect the main broad emission
lines in a typical BLAGN spectrum. 

\section{Spectral classification}
\label{Catalog}
  \begin{figure}
   \centering
   \resizebox{\linewidth}{!}{\includegraphics{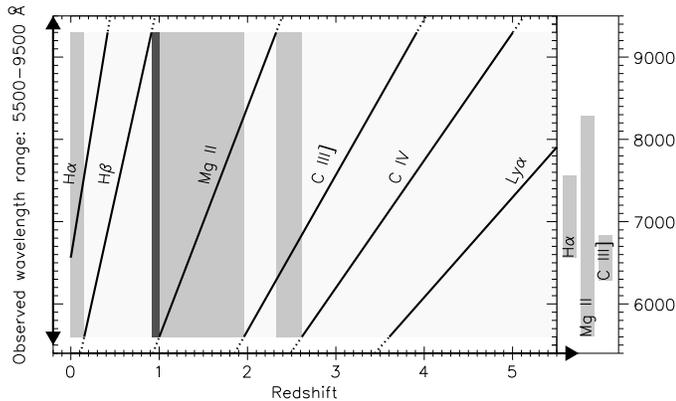}}
   \caption{\textbf{Visibility in the wavelength range of our data of
   AGN broad emission lines along redshift:} 
   The thick lines trace the observed wavelength as a function of
   redshift of the main AGN broad emission lines that we expect to
   detect in our spectra.
   The objects in which we detect only a single broad emission line
   are expected to be in the redshift ranges shown in light gray.
    We expect to miss BLAGN lying in the narrow redshift range filled in dark
   gray since no strong broad emission line is visible at that
   redshift within our observed wavelength window. 
   On the right side are reported the wavelength ranges in which
   each broad emission line can be observed as isolated.
   H$\beta$ and Ly$\alpha$ are not reported there since they are not concerned
   by the redshift degeneracy.
   }
    \label{EL_diag}%
    \end{figure}
%--------------------------------------------------

\subsection{The VVDS spectroscopic catalog}
The process of classification and redshift determination of the reduced
VVDS spectra is described in \cite{LeFevre2005}. As an integrated part
of this process all objects with broad emission lines are identified
and flagged.
When only narrow emission lines are present, it is not easy to
recognize AGN activity (type-2 AGN) from starburst activity, 
especially in spectra with relatively low S/N ratio 
and limited wavelength range. 
For this reason  we have chosen not to include narrow line AGN in our
present catalog. 
This population of objects will be presented in forthcoming
publications.

\subsection{The BLAGN catalog}
\label{BLAGNCatalog}
Our BLAGN catalog consists of the sub-sample of
objects in the VVDS spectroscopic catalog identified purely on the
basis of the presence of one or more broad emission lines.
We have selected objects with broad lines full width half maximum (FWHM)
larger than $1000$ ${\rm km\,s^{-1}}$ , a secure threshold compared to our
spectral resolution.   
We make no distinction between Seyfert galaxies and
QSOs based on the absolute magnitude or morphology of the
objects. 

The final sample of BLAGN spectra selected from the VVDS survey
consists of 56 objects from the wide fields (``the wide BLAGN sample'') 
and 74 objects from the deep fields ("the deep BLAGN sample").
The total number of 130 BLAGN corresponds to about 0.7\% of the objects 
in the total VVDS database with a measured redshift.
Not all of the 130 BLAGN selected have a secure redshift. 
In some cases, we have two or more possible redshifts for a given objects
(`redshift degeneracy'). 

\subsubsection{Redshift degeneracy}

The exact list of emission lines we expect to detect in any given
spectrum is a function of its S/N. In addition,
because of the limited wavelength coverage of the observations, there
are some redshift intervals where only a single broad emission
line can be detected in our spectra (see Figure \ref{EL_diag}). 
In these cases, if
no other features (e.g. narrow emission lines, absorption systems,
Lyman forest) are present, several interpretations of the broad line
may be allowed, thereby leading to a degeneracy of solutions for the
redshift (see an example in Figure \ref{showspectra}).

In order to treat redshift degeneracy in a consistent way, we have taken
the SDSS composite spectrum of BLAGN \citep{VandenBerk2001}, 
convolved it to our resolution, and added noise. 
For each line we then determined the S/N which would
provide a 3.5 $\sigma$ detection of the line.
Following this procedure we have found that down to a S/N of 4 per resolution
element, we are able to detect and recognize  
as broad the following lines: H$\alpha$ 6563, H$\beta$
4861, MgII 2799, C\,III] 1909, C\,IV 1549 and Ly$\alpha$ 1216.

The visibility windows of these lines within the VVDS observed spectral
range are shown in Figure \ref{EL_diag} as a function of redshift.
The redshift ranges where we expect to observe a single broad emission line
(in gray) are: [0 - 0.13], [1.0 - 1.9] and [2.4 - 2.6].
  
Since we always expect to detect the narrow-emission-line
doublet [OIII] 4959 5007 close to H$\beta$, we  consider the detection of
H$\beta$ to securely identify the redshift. 
The same applies to $Ly\alpha$ which can be identified from the Lyman
forest. 

Consequently, for an object with a single broad line, we 
have one to three potential redshifts.
In addition, a small gap is present at redshift
$\sim$0.95 (dark gray in Figure \ref{EL_diag}) where no broad emission line is visible
within our observed wavelength window.

A small fraction of spectra ($\sim 15 \%$) fall outside this general
scheme, due to a lower S/N or to technical limitations such as strong
fringing. For this reason some objects have up to four possible redshifts. 

Because of these limitations, a total of 42 BLAGN in our original sample were
flagged as having a degenerate redshift (Flag 19, as defined section
\ref{section:flag}). 
To solve this redshift degeneracy we have first looked for the objects already
observed in other spectroscopic surveys in the same areas.
From the CDFS optical spectroscopic data we found the redshifts of 3 of the
objects \citep{Szokoly2004}.  
Furthermore, additional observations have been obtained with FORS1 on the VLT
in March 2004 and November 2005 to extend the spectral coverage down to 3800 \AA\ .
From these observations we have been able to solve the redshift degeneracy and
find a secure redshift for 25 objects.  

The fraction of BLAGN remaining with a degenerate redshift corresponds
to $\sim 10$ \% of the total BLAGN sample.

\subsubsection{\label{section:flag}Identification and redshift quality flags}

Based on the above analysis, each entry of our BLAGN catalog has been
assigned a quality flag according to the following criteria
(these flags have been adapted from the VVDS galaxy flag definition): 
\begin{description}
\item[Flag 14:] BLAGN with secure redshift measurement
  (i.e.: two lines or more); 
\item[Flag 13:]  BLAGN with only one line detection
      but with the redshift secured from other informations
      such as:
      \begin{itemize}
      \item no other possible identification for the given
        wavelength range;
      \item an intervening absorption line system is detected and this leaves only
      one redshift solution for the BLAGN;
      \item strong iron features, so called 'bumps', are detected;
      \item a second line is marginally detected;
      \end{itemize}
    \item[Flag 19:] BLAGN with only one detected line and
      more than one possible redshift;
    \item[Flag 12:] Object with secure redshift but
      for which the broad line nature is
      uncertain due to the low spectral resolution, often coupled with
      low S/N or to technical limitations  (i.e. it could be a narrow line
      object);  
    \item[Flag 11:]Both the redshift and the broad line nature of 
      the target are uncertain.
\end{description}

%__________________________________________________ Figure quality flags 
\begin{figure}
  \centering
     \resizebox{9cm}{!}{\includegraphics{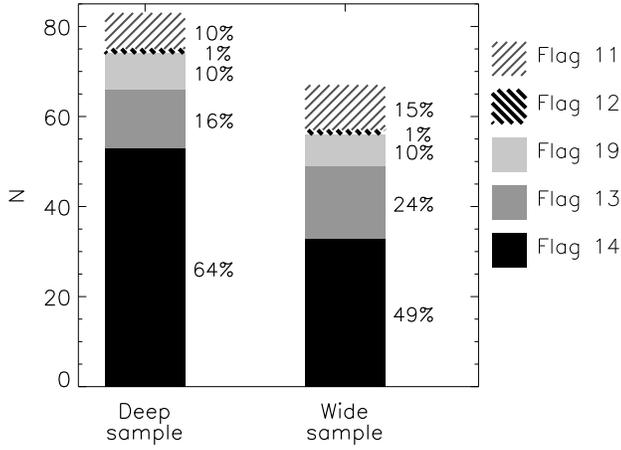}}
  \caption{AGN quality flags statistics.}
  \label{flagstat}
\end{figure}

The statistics of the data quality according to these flags is
summarized in Figure \ref{flagstat}.  
Objects with uncertain broad emission lines (flags 11 and 12), 
corresponding to $\sim$(10-15)\% of our survey objects, are
excluded from the present analysis, since we do not want to be
contaminated by normal galaxies and we do not know which
proportion of these objects are really broad-line AGN. 
The final sample of BLAGN with secure classification (flag 14, 13 and 19), 
used in the statistical analysis, contains 130 objects.
They are presented in Appendix \ref{tab}: Table \ref{tab:AB} gives the list of
BLAGN in our wide and deep samples for which we have a secure 
redshift, while Table \ref{tab:C} lists the BLAGN for which we still
have a redshift degeneracy (flag 19).

\begin{figure}
     \centering
     \resizebox{\linewidth}{!}{\includegraphics{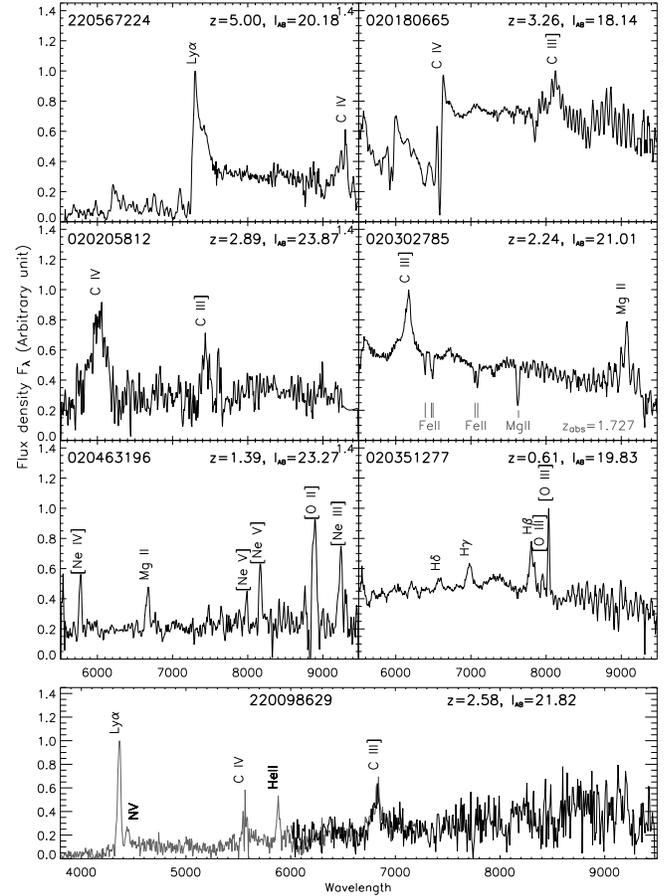}}
     \caption{Examples of BLAGN spectra across our magnitude and
     redshift ranges. Spectra of AGN with $I_{AB} > 23$ have been rebinned to
     5 pixels.}
     \label{showspectra}
   \end{figure}

Figure \ref{showspectra} presents some examples of spectra across
our magnitude and redshift range. 
In particular object 220567224 is the highest redshift BLAGN discovered 
with this survey; object 020180665 is an example of 
a broad-absorption-line (BAL) spectrum, while object 020302785 is instead a
BLAGN at z=2.24 in which there are evident absorption features that can be
well fitted by metal absorptions of a system at z$\sim$ 1.727. 
Object 020463196 is an example of AGN in which strong [NeIII], [NeIV] and
[NeV] lines are visible and object 220098629 is an example of
flag 19 for which the redshift was resolved with additional observations
obtained with F0RS1.

%__________________________________________________ Figure showspectra
%-------------------------------------------------

 \section{\label{SF}Selection function}
To study the statistical properties of the VVDS BLAGN sample we need to
correct for the BLAGN which have not been spectroscopically observed and
for the BLAGN which were not correctly identified from their spectrum.
This correction is performed applying to our BLAGN two statistical weights, 
 $w^{TSR}$ and $w^{SSR}$ defined hereafter following \citet{Ilbert2005}.

    \subsection{\label{TSR}Treatment of non-targeted BLAGN: $w^{TSR}$}
The \textit{Target Sampling Rate} (hereafter TSR) is the fraction of objects
in the photometric catalog inside our targeted area which have been
spectroscopically observed.  As already mentioned in Section \ref{selection},
the TSR is a function of the projected size, $x-radius$, of the objects along
the slits.  Objects with a small x-radius are slightly favored by the
selection process.  
 To balance for non-targeted BLAGN we apply to our BLAGN the weight $w^{TSR} =
\frac{1}{TSR}$.

    \subsection{Treatment of misclassified AGN: $w^{SSR}$}
\begin{figure}
     \centering
     \resizebox{\linewidth}{!}{\includegraphics{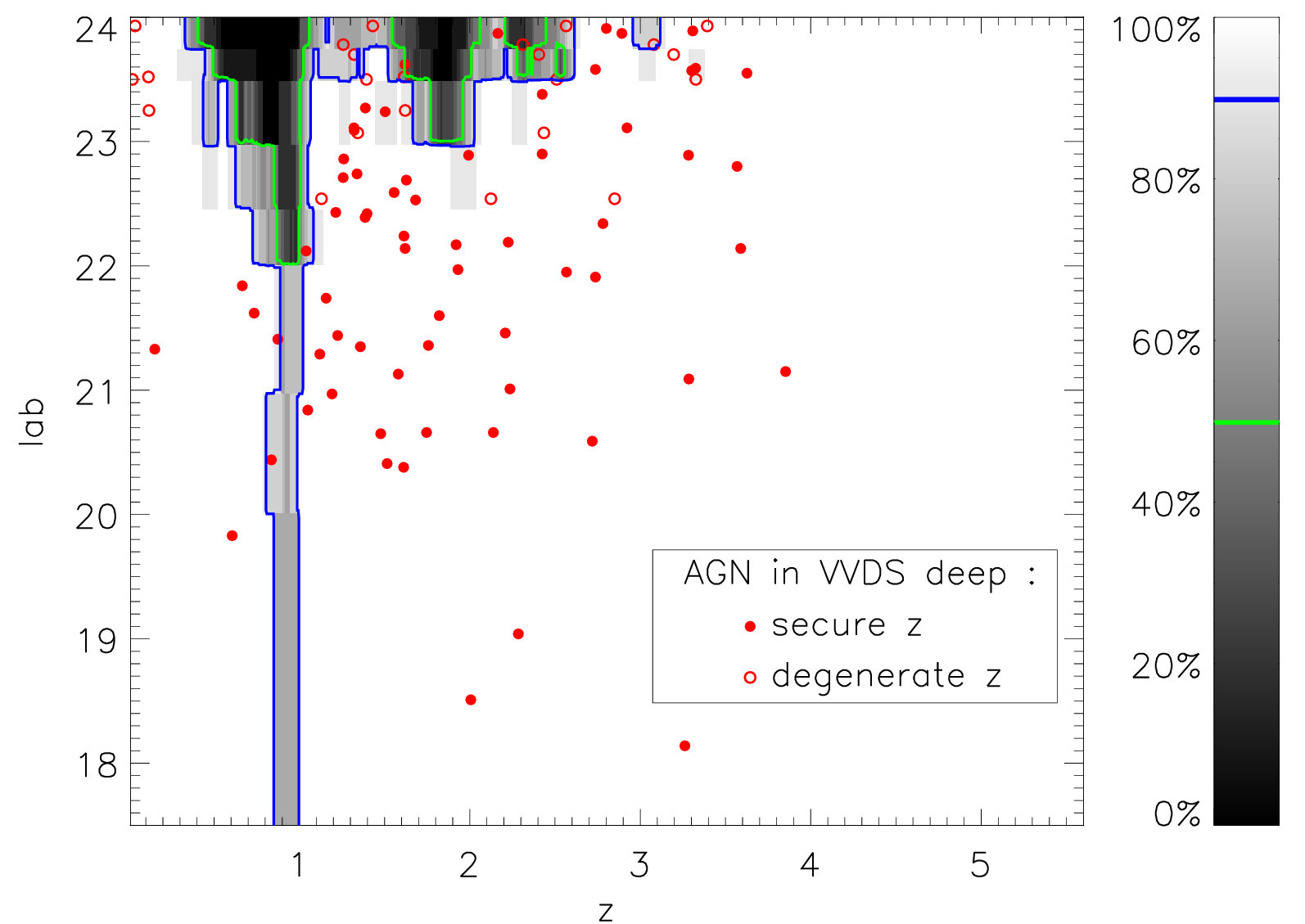}}
     \resizebox{\linewidth}{!}{\includegraphics{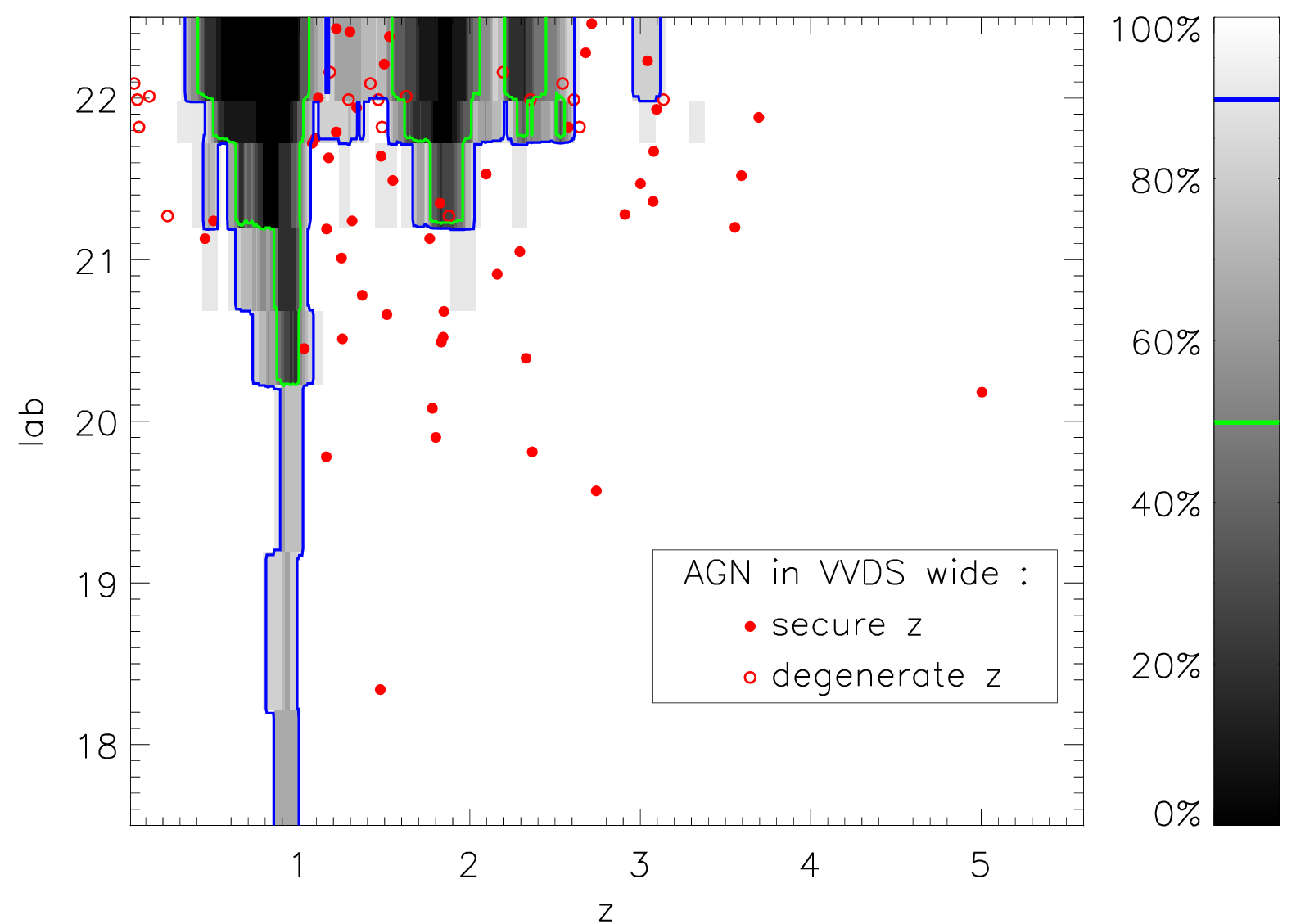}}
     \caption{\label{fig:SSR} \textit{Spectroscopic Succes Rate} for BLAGN 
     in the VVDS deep (top) and wide (bottom) sample as a function of redshift
     and apparent magnitude.
     Contour lines are drawn for 90\% and 50\% of identification success
     rate.
     The location of BLAGN with a secure redshift is reported with
     filled dots. 
     BLAGN with a degenerate redshift are plotted at their different redshift
     solutions with open circles.}  
   \end{figure}

The \textit{Spectroscopic  Success Rate} (SSR) is the probability of a
  spectroscopically targeted object to be securely identified.
It is a complex function of the BLAGN redshift, apparent magnitude and
intrinsic spectral energy distribution. In order to evaluate the
SSR, we simulated 20 Vimos pointings, for a total of 2745 spectra. The
simulations incorporate the major instrumental effects of Vimos
observations, including sensitivity curve, fringing in the redmost part
of the spectra, sky background and contamination by zeroth-orders. Each
spectrum has been simulated using the SDSS composite spectrum as a
template. Redshifts and magnitude were drawn at random in the ranges
$0<z\le 5$ and $17.5\le I \le 24$ respectively. The spectra were then
analyzed and classified in the same way as the real spectra, except that we
did not try to resolve cases where a single line was present; we merely
checked whether the BLAGN nature of the object could be detected in the
spectrum or not. 
Figure \ref{fig:SSR} presents the resulting SSR as a function of
  redshift and apparent $I_{AB}$ magnitude in the deep and wide fields.
To correct for missed BLAGN we apply to our objects the weight $w^{SSR}
  =\frac{1}{SSR}$. 

 Objects for which the broad-line equivalent width (EW) is larger than in the
  SDSS composite spectrum will be over-estimated by our method and, on the
  contrary, objects with weaker broad emission lines will be under-estimated.  
We investigated in particular the differences in the SSR which are
  expected for AGN with the emission lines properties of our composite
  spectrum (see Section \ref{sec:composite} and Table \ref{tab:lines}). We
  found this differences to be small.
  
%__________________________________________________ Figure SSR
%--------------------------------------------------

Correction for our global selection function is obtained by applying to each
of our objects the product of these two weights, i.e. $w^{TSR} \times w^{SSR}$.

 \section{Counts}
  \label{Counts}
%__________________________________________________ Figure surf. dens.
\begin{figure}
  \centering
  \resizebox{9cm}{!}{\includegraphics{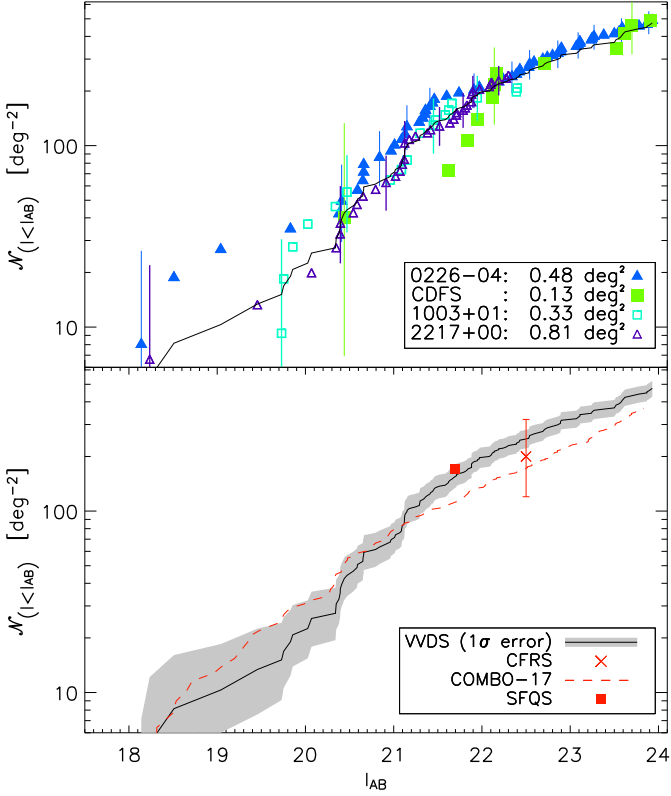}}
  \caption{\label{fig:dens} BLAGN cumulative surface density in the VVDS
   sample. Error bars are the one sigma Poissonian errors. 
  Top panel: Observed surface density in the individual fields and
   in the total sample (solid line). For readability error bars are reported
  each 5 data points.
  Bottom panel: Comparison with CFRS and COMBO-17 surveys.}
\end{figure}
%--------------------------------------------------
%_________________________________________________ Table surf. dens.
\begin{table}
      \caption[]{\label{tab:counts}
BLAGN number counts as function of $I_{AB}$ magnitude. 
N is the actual number of BLAGN in the VVDS survey while $\mathcal{N}\left(\le
  I_{AB} \right)$ is the cumulative surface density of BLAGN (objects / square
degree) corrected from incompleteness.}  
\begin{center}
\begin{tabular}{rrr}
   \hline \hline
$I_{AB}$ & N & $\mathcal{N}\left(\le I_{AB} \right)$ \\
\hline
 19 &   3 &  10 \\
 20 &   9 &  22 \\
 21 &  29 &  71 \\
 22 &  76 & 196 \\
 23 & 108 & 327 \\
 24 & 130 & 472 \\
\hline \hline
\end{tabular}
\end{center}
\end{table}
%--------------------------------------------------

Since the objects to be observed spectroscopically in the VVDS are selected on
the  basis of their I$_{AB}$ band magnitude, we have computed the BLAGN
cumulative surface density in this band with the following algorithm:
$$
\mathcal{N}\left(\le I_{AB} \right) = 
\frac{1}{\mathcal{A}} 
\sum_{i,I_{AB,i} \le I_{AB}} w_i^{TSR} w_i^{SSR}
$$
where $\mathcal{A}$ is the geometrical area targeted by the
  spectroscopic survey, i.e.  the area covered by the different VIMOS
  fields of view corrected for the area masked in our photometric catalogs;
and, where the sum is running over all BLAGN $i$ with magnitude $I_i$ less or equal to $I_{AB}$. 

The upper panel of Figure \ref{fig:dens} presents the cumulative surface density derived in
the VVDS individual fields  with the corresponding Poissonian error bars \citep{Gehrels1986}.
Although some differences in the counts in different fields are
visible, the surface densities measured in all our fields at $I_{AB} =22.5$,
corresponding to the magnitude limit of the wide fields, are all consistent,
within the errors, with each other.
We conclude that the error induced by cosmic variance is smaller than our Poissonian noise. 
We compute a single $log \mathcal{N}-I_{AB}$ curve by merging the data
from the four fields into a single coherent sample \citep{AvniBahcall}. 
%__________________________________________________ Figure differential surf. dens.
\begin{figure}
  \centering
  \resizebox{9cm}{!}{\includegraphics{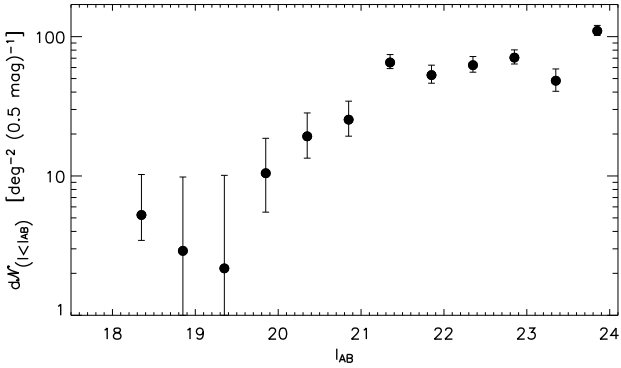}}
  \caption{\label{fig:ddens} BLAGN differential surface density in the VVDS
   sample. Error bars are the one sigma Poissonian errors. }
\end{figure}
%--------------------------------------------------
At the limit of our deep fields ($I_{AB}< 24.0$) our measured 
surface density of BLAGN is $\sim 472 \pm 48$ per square degree and the
correction applied for misclassified BLAGN corresponds to $\sim 10\%$ of this
value. 
The differential surface density
  computed for the total sample is shown in Figure \ref{fig:ddens}. In
  this plot the different points and associated errors are independent from
  each other. This figure suggests a significant turn-over in the slope of the
  counts at $I_{AB} \sim 21.5$. At $I_{AB} \gtrsim  21.5$ the differential number
  counts are still increasing with magnitude, but at a much lower rate than at
  brighter magnitudes. 
  This effect is seen, although less clearly, in the continuous
  flattening with magnitude of the integral number counts in Figure
  \ref{fig:dens}.

Comparison with other spectroscopic surveys is not straightforward
since most of the optical QSO surveys are based on much shallower
B band flux limited samples and most of them are not complete over the entire
redshift range due to their different selection criteria.
The CFRS is an exception with essentially the
same selection function as the VVDS (i.e. a flux limited sample in the I band
with no color nor morphological selection). 
Although it contains only 6 BLAGN at a magnitude limit of $I_{AB} < 22.5$
\citep[see][]{Schade1996}, 
the surface density resulting from these objects  is in excellent agreement 
with our current measurement (see Figure \ref{fig:dens}, lower panel).

The surface density derived from the recent SDSS Faint Quasar Survey
\citep[SFQS][]{Jiang2006} is also reported in Figure \ref{fig:dens}, lower
panel. 
This survey selects AGN over $0 < z < 5$ up to $g=22.5$. The surface density,
corrected for completeness, is computed in the $g$ band (Jiang, private
communication) and approximately translated to the $I_{AB}$ band assuming the
mean color term observed in our sample: $< I_{AB} - g > = - 0.7$. 
%Although this simple comparison becomes invalid for objects with $z<2.3$ 
The VVDS and SFQS number counts at $I_{AB} \sim 21.8$ are in good agreement.

With a limiting magnitude similar to ours the only other comparable sample is
the COMBO-17 sample \citep[see][]{Wolf2003}, which is however based on
photometric selection of AGN, with a small fraction of spectroscopic
confirmation. 
For comparison with our sample, we have obtained the complete COMBO-17 sample
cumulative counts, without redshift restriction (Wisotzki, private
communication). 
Since this sample is selected in R-band, we apply a global color term of
$-0.16$ magnitude to translate their surface density to our $I_{AB}$ reference
system. This color term corresponds to the mean $I_{AB} - R_{vega}$ color
observed for our objects. In addition, a dozen of BLAGN are observed in both
VVDS and COMBO-17 sample. The mean color term of these objects is measured to
be $-0.12$.  
The result is presented in the lower panel of Figure \ref{fig:dens}.
Our number counts are statistically consistent with COMBO-17 for magnitudes
brighter than 21.5. At fainter magnitude our counts are systematically higher
by $\sim 30-40$\%. 
This can be explained by incompleteness in the COMBO-17 at
low redshift in the regime where the host galaxy contamination becomes
non-negligible. This is the reason why the published version of the COMBO-17
AGN sample was restricted to $z > 1.2$.
When restricting our sample to $z > 1.2$ and compute our surface
  density in the R band, we find $\mathcal{N}(R \le 24) = 340 \pm 47$ which
  compares well with the published result of COMBO-17 sample $\mathcal{N}(R
  \le 24) = 337$.

\section{Redshift distribution}
\label{Zdistrib}
%__________________________________________________ Figure histz
   \begin{figure}
     \centering
     \resizebox{\linewidth}{!}{\includegraphics{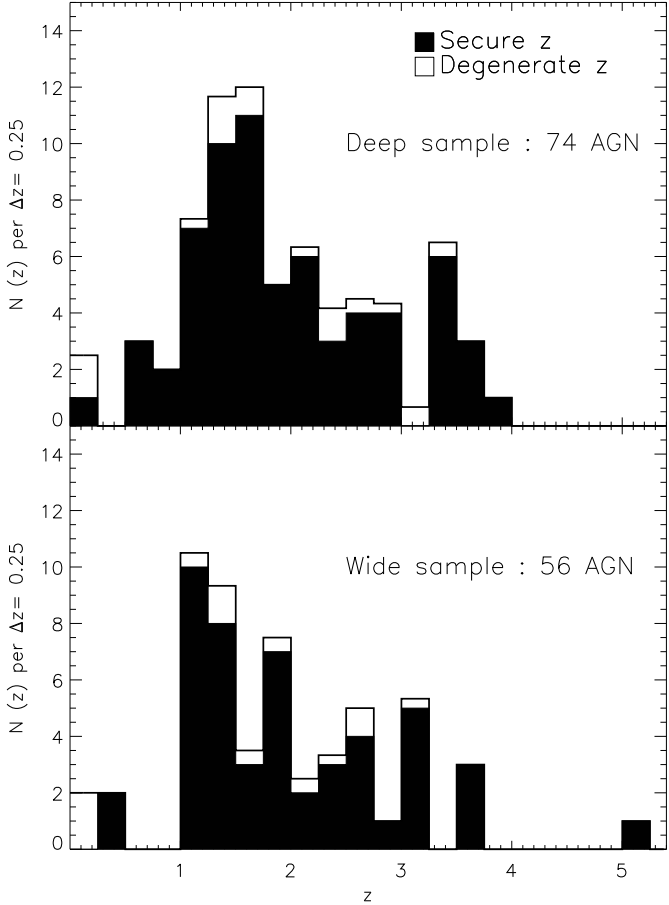}}
     \caption{Redshift distribution of the VVDS BLAGN sample.
       The shaded histogram corresponds to BLAGN with a secure
       redshift (flags 13 and 14), while the unshaded histogram
       includes also flag 19 BLAGN, distributed with an equal
       weight over the different redshift alternatives.}
     \label{histz}
   \end{figure}
%--------------------------------------------------
The redshift distributions of our wide and deep samples are presented in
Figure \ref{histz}.  BLAGN with degenerate redshift solution (flag 19) have
been distributed with equal weights in the different possible redshifts.
The absence of objects between redshift 0.5 and 1 in the wide fields can be
attributed to our low efficiency for selecting objects at this redshift (see
Figure \ref{fig:SSR}).
The main uncertainty resulting from BLAGN with a degenerate redshift
concerns the fraction of BLAGN at low redshift.

The fraction of $z > 3$ objects is $\sim$ 17\% in the wide sample and $\sim$
15\% in the deep sample.
The mean redshift is $ \sim 1.8$ in both the wide and deep samples.
Considering all flag 19 BLAGN successively at their lower and  higher
redshift solution we find that the mean redshift of both deep and wide sample
lies between 1.5 and 2.1.
This shows that, by pushing our limit in magnitude from the wide to the deep
survey, we  are not increasing our mean redshift but rather exploring the
fainter part of the luminosity function at all redshifts. 

%__________________________________________________ Figure mag=f(Z)
   \begin{figure}
     \centering
     \resizebox{\linewidth}{!}{\includegraphics{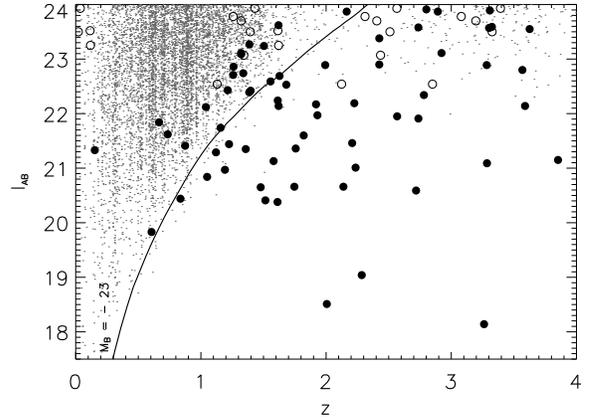}}
     \caption{Apparent $I_{AB}$ magnitude as a function of redshift in the
       VVDS deep sample. Galaxies are plotted with points, BLAGN
       with secure redshift are plotted with filled dots and BLAGN with
       degenerate redshift are presented at all their redshift
       solutions with empty circles.
     The thick line gives the apparent magnitude of a BLAGN template of
       absolute magnitude $M_{B} = -23$ and therefore corresponds to
       the transition between the QSOs and Seyfert galaxies.}
     \label{magz}
   \end{figure}
%--------------------------------------------------

Figure \ref{magz} shows the apparent magnitude versus redshift
distribution of BLAGN and galaxies in the deep sample. 
We see that some of our BLAGN are sampling the faint magnitudes, well
inside the galaxy luminosity function.
We find that $\sim 35$\% of the objects in the deep sample are Seyfert
galaxies, rather than QSOs, having absolute magnitudes fainter than
$M_{B}=-23$.  

\section{Morphology and color properties}
\label{Morphcol}
The fundamental property of our survey is that it does not suffer neither 
from morphological nor from color selection biases which are present in
most  optical surveys.
Therefore, using the VVDS deep photometric data, we are able, a posteriori, to
quantitatively estimate the fraction of AGN missed by these standard
selection criteria. 

\subsection{Morphological analysis}
\label{sec:morph}
The use of a morphological selection of point-like objects causes the loss of
BLAGN candidates for two main reasons.
The first is due to the fact that a reliable separation between
point-like and extended sources from ground-based data images is
possible only for relatively bright objects. The second is linked to
an intrinsic property of BLAGN:  
at relatively low z and low intrinsic luminosity, BLAGN
may appear extended or at least slightly resolved because of
significant contribution from the host galaxy. This 
effect can introduce a redshift-dependent bias in morphologically selected
AGN samples, especially at faint magnitudes. 

We have conducted a morphological analysis of our BLAGN sample through
the study of various shape parameters, which define a characteristic
radius weighted by the light distribution function \citep{Kron1980}. 
In particular, for this analysis we used the \texttt{SExtractor}
flux-radius parameter, denoted as $r_{2}$, which measures
approximately the radius which encloses half of the object's total
light.  
This parameter is used to identify point-like sources, because, for
unresolved objects, it is directly related to the width of the point
spread function and it is independent of luminosity (for non-saturated
objects).   
In the plane $r_{2}$, measured in I-band, versus I-magnitude, we can
distinguish two different classes of objects: unresolved  
sources that occupy a well defined strip of the plane at small and
approximately constant values of $r_2$  and resolved objects for which
$r_2$ varies with magnitude.  

This analysis is restricted to F02, F10 and F22 fields. We exclude the CDFS 
for which this parameter is not measured.
Since our data, covering large areas in different sky regions, have
been taken with different seeing conditions, the stellar locus
resulting from the use of the measured $r_{2}$ values is significantly
broadened, thus making a clear separation between point-like and
extended objects difficult. 
To account for this, we used an ``adaptive'' classification technique
following \citet{McCracken2003}: after dividing the field
into many sub-areas, mainly following the pattern corresponding to
different pointings, we have normalized  
the flux radius of each object to the local $r_2$ mode, computed for
all the objects within each sub-area. 
%__________________________________________________ Figure grafico_finale_3.5sigma.ps
\begin{figure}[htb]                                                   
\begin{center}                                                      
\includegraphics[height=9.0cm,width=9.0cm]{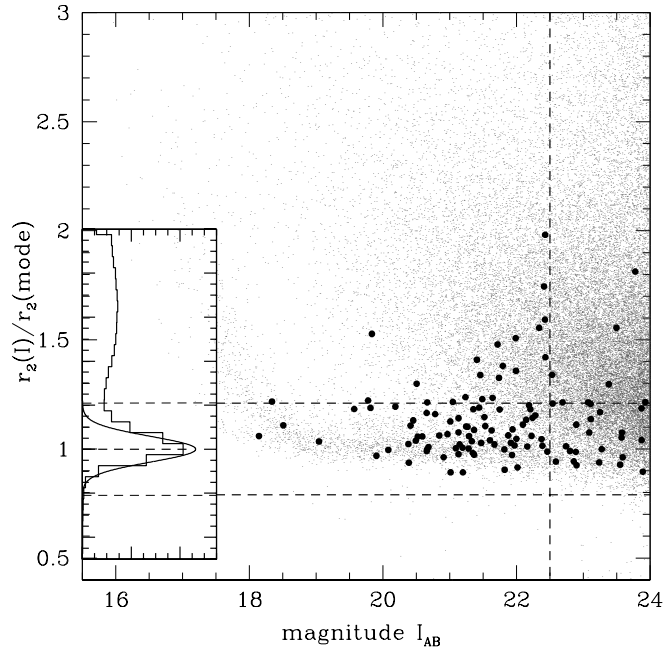}
\caption{Normalized $r_{2}(I)$ parameter versus $I_{AB}$ magnitude.
The two horizontal lines correspond to the range  in $r_{2}(I)$ we
have adopted for our morphological classification of point-like
sources. Large circles are the spectroscopically confirmed BLAGN. The inset
shows the $r_{2}$ distribution of objects with $21.0 < I_{AB} < 22.0$, fitted
with a Gaussian with $\sigma$ = 0.06.} 
\label{fig:morph}                                                     
\end{center}                                 
\end{figure}
%--------------------------------------------------
Figure \ref{fig:morph} shows this normalized $r_{2}$(I) parameter versus
I-magnitude.  

The use of the normalized $r_{2}$(I) allows us to improve our
ability to distinguish point-like from extended objects. 
As shown in Figure \ref{fig:morph}, this separation appears to be
feasible up  to $I_{AB} \sim 22.5$ (without normalization  
the corresponding limiting magnitude was $I_{AB} \sim 21.5$).
Beyond this magnitude ($I_{AB} \sim 22.5$),  the number of galaxies with
small $r_2(I)$ parameter increases and no reliable selection of
point-like sources is possible. 

The two horizontal lines correspond to the range in $r_{2}(I)$ we have
adopted for our morphological classification of point-like sources.
The upper limit in this range corresponds to about 3.5$\sigma$ of the
Gaussian fit of the $r_{2}$ normalized distribution of point-like
sources. 
As shown by the inset in the left part of Figure \ref{fig:morph}, this
somewhat conservative choice makes sure that the vast majority of point-like
sources are really classified as point-like, allowing some contamination from
extended objects, especially at faint magnitude. Efficiency in classifying point-like
sources has been tested on the spectroscopically confirmed stars in
the magnitude range $18.5 \leq I_{AB} \leq 22.5$ (the bright limit has
been  set in order to exclude saturated objects (see Figure
\ref{fig:morph})). We find that 95\% of the stars with spectroscopic
flag $\geq$ 3 \citep[see][]{LeFevre2005} are correctly
classified as point-like sources.

Figure \ref{fig:morph} also shows the location in this plane of the
spectroscopically confirmed BLAGN. 
In this analysis we consider only BLAGN brighter than $I_{AB}$ = 22.5 (90
objects); within this limit, 77\% of the BLAGN are classified as point-like, while
21 of them (23\%) are classified as extended. 
This percentage is significantly higher than that of morphologically
misclassified stars (see above). 

%__________________________________________________ Figure z_distrib_3.5
\begin{figure}[htb]                                                   
\begin{center}                                                      
\includegraphics[height=5.0cm,width=8.0cm]{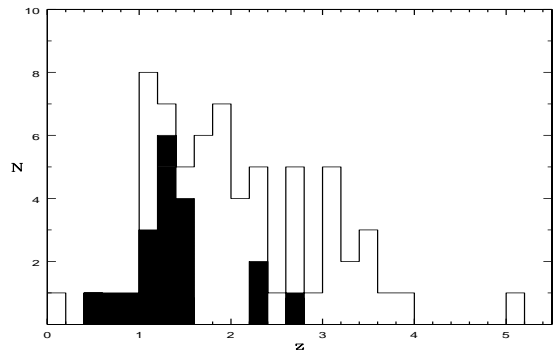}
\caption{Redshift distribution of BLAGN with secure redshift and $I_{AB}<22.5$.
 The unshaded histogram show the objects classified as point-like while the
 filled histogram shows the objects classified as extended.}
\label{fig:zdistr}                                                  
\end{center}                                 
\end{figure}
%--------------------------------------------------

Nineteen of these 21 extended BLAGN, have a secure redshift measurement,
while the other 2 are BLAGN with two or more possible values for z.
 Figure \ref{fig:zdistr} shows the redshift distribution of BLAGN with
 secure redshift and $I_{AB} \leq 22.5$, morphologically classified as
 extended compared to those classified as point-like.  
The two redshift distributions are significantly different (the
probability that they are drawn from the same distribution, as
estimated from a Kolmogorov-Smirnov (KS) test, is smaller than $\sim 4 \times
10^{-4}$), with the extended BLAGN having on average a lower redshift. 
The Figure show that sixteen out of nineteen of these BLAGN have $z < 1.6$ 
and in this redshift range they constitute $\sim$ 42\% of the sample.  

\subsection{Color analysis}
The classical way to create an optical BLAGN sample is to preselect
candidates from a photometric catalog requiring that the objects are
point-like and have blue colors. 
This selection criterion allows to select BLAGN candidates with an efficiency
above 50\% \citep[see][]{Croom2004}, at least for objects brighter than B$\sim$21.  
This jump in efficiency --from less than 1\% in a flux-limited sample
without preselection (see end of Section \ref{BLAGNCatalog}) 
to more than 50\% with preselection -- makes it possible to
construct very large catalogs of several tens of thousands of BLAGN, such as
the 2QZ.  

The drawbacks of this selection criterion are: 
\begin{description}
\item[(a)]{Its high redshift limit arises at about the same
  redshift where the cutoff of BLAGN space density is observed, i.e $z
  \sim 2.3$;}
\item[(b)]{It excludes objects at low redshift, for which the
  contribution of the host galaxy is detected and resolved, thus
  introducing a bias especially toward the faint end of the BLAGN luminosity
  function.}  
\end{description}

To overcome this problem, various surveys adopted more complex
selection algorithms based on multicolor analysis, estimating
for each of them different levels of
incompleteness. The SDSS QSO survey is the largest example of this
kind of survey. Its completeness function is investigated in detail in 
\cite{VandenBerk2005}.
%--------------------------------------------------
\begin{figure*}
 \resizebox{\linewidth}{!}{\includegraphics{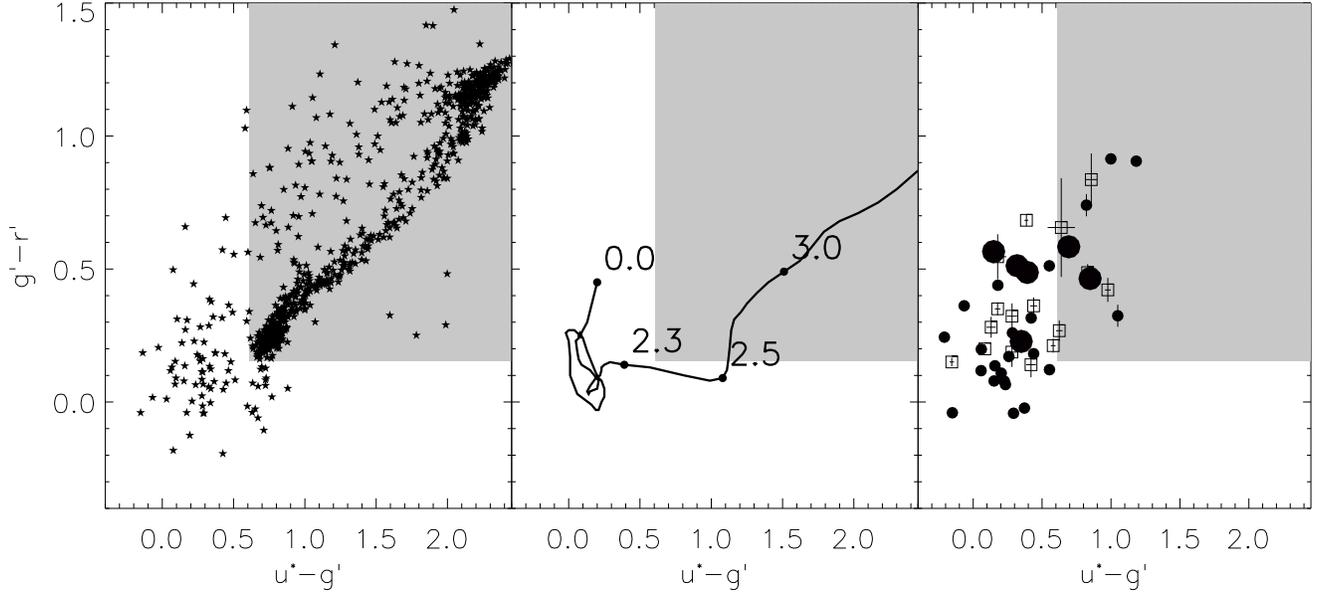}}
\caption{\label{fig:colorselect} $u^*-g'$, $g'-r'$ color diagram. 
  Left panel: location of the point-like objects in our photometric
  catalog. An exclusion area, reported in gray, is set to exclude
  most of main sequence stars.
  Middle panel: color evolutionary track with redshift of a
  BLAGN. The template considered here is the SDSS composite spectrum.
  Right panel: location of AGN with $z<2.3$. AGN which could not be
  morphologically classified ($I_{AB} > 22.5$) are plotted with open boxes. AGN
  classified as extended are big dots, while AGN classified as point-like are
  small dots.}
\end{figure*}
%--------------------------------------------------

We investigate here how a simple utraviolet excess preselection technique based
on three color bands would have applied to our sample. 
For this purpose we used the deep photometry obtained in the frame of the
CFHTLS and we plotted our point-like objects in the color plane $u^*-g'$,$g'-r'$,
as shown Figure \ref{fig:colorselect}, left panel. 
We could then set a color selection criterion meant to exclude most of the
main sequence stars: (a) $u^*-g'< 0.62$ ; (b) $g'-r' < 0.16$.
The exclusion area is reported in gray in Figure \ref{fig:colorselect}. 
Most of the objects in the exclusion area, ouside the locus delineated by the
main sequence stars are compact galaxies classified as point-like objects.

Following the evolutionary track of a QSO template (here the SDSS composite
spectrum, \citealt{VandenBerk2001}) in this color plane, we can expect this
selection criterion to be efficient up to $z\sim 2.3 - 2.5$. 

We estimate now which fraction of the optical AGN population with $z<2.3$
would have been missed by this selection criterion. 
Our AGN sample with $z<2.3$ and photometric information in the $u^*$,$g'$ and $r'$
band is reported in this plane (Figure \ref{fig:colorselect}, right panel).
We find that $\sim 25 \%$ of the population of our optical BLAGN would have
been excluded by this color selection.
If we restrict now our analysis to a limiting magnitude of $I_{AB} < 22.5$,
we find that $\sim 35 \%$ of our AGN do not fulfill this color selection
criterion combined with a morphological selection of point like objects.
As expected, at higher redshift this selection criterion becomes
  inefficient, with 75\% of our $z > 2.3$ objects excluded by this criterion.
We would like to emphasize that these completeness rates are found for
our sample which is two to three magnitude deeper than current major samples
and do not necessary apply to existing brighter samples.

%__________________________________________________ Figure BIz
   \begin{figure}
     \centering
     \resizebox{\linewidth}{!}{\includegraphics{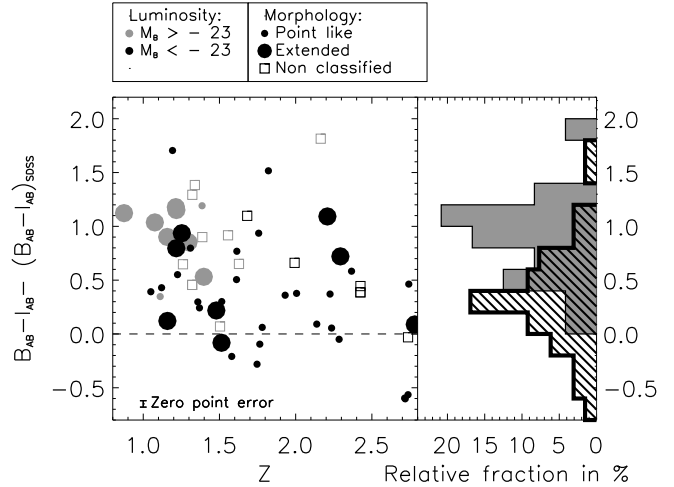}}
     \caption{Difference as a function of redshift between the observed
     $B_{AB}-I_{AB}$ colors and the color expected for the SDSS composite
       spectrum. 
       Black dots correspond to QSOs, while the
       gray dots correspond to Seyfert galaxies. 
       The $B_{AB}-I_{AB}$ color histograms 
       of these two populations are shown on the right.}
     \label{fig:BIz}
   \end{figure}
%--------------------------------------------------

Figure \ref{fig:BIz} presents the difference as a function of redshift between
the $B_{AB}-I_{AB}$ colors of our BLAGN sample and the color expected for the
SDSS composite spectrum for $z < 2.8$. 
The dispersion in color is large and most of our objects have a significantly
redder color than that expected from the SDSS composite spectrum.
 Moreover, the faintest objects appear to have on average redder
colors than the brightest ones. A KS test applied on the color of the
Seyfert galaxies ($M_B > -23$) and of the QSOs ($M_B < -23$) indicates a
probability $\sim 1 \times 10^{-3}$ that these objects have the same color
distribution. 

This could be explained in different ways:
(a) the contamination of the host galaxy is significantly reddening the
measured colors of faint AGN; 
(b) BLAGN are intrinsically redder when they are faint; 
(c) the reddest color are due to the presence of dust.
With the present analysis we can not exclude or confirm hypotheses (b) and
(c). 
However, we observe that for $\sim 50-60$ \% of our reddest AGN (objects with
$(B_{AB}-I_{AB})-(B_{AB}-I_{AB})_{SDSS} > 0.8$) we detect the extended
component of the BLAGN host galaxy in I band. 
 In this redshift range the contribution from the host galaxies is in fact 
 expected to be more significant in the I band (corresponding to near UV-B
 rest-frame bands) than in the B band (corresponding to the far U rest-frame band).
We therefore believe that host galaxy contamination is contributing to give a
redder color to our BLAGN. 

From this morphological and color analysis we conclude that classical
optical preselection techniques are significantly under-sampling
the overall BLAGN population in deep samples. This effect
should be fully taken into account when deriving BLAGN volume 
densities or related quantities.
 
\section{Composite Spectrum}
\label{sec:composite}
A rest-frame composite spectrum was generated  according to the method
described in \cite{Francis1991}.
The spectra were first blue-shifted in their rest frame and rebinned
to 1 \AA .  
The lowest redshift spectrum was taken as a starting point to build
the long wavelength part of the composite.  
The spectra were then taken in order of increasing redshift and
for each of them the spectral range in common with the composite was
determined. The spectrum was rescaled to have the same average as the
composite in this common spectral range. It was then added to the
composite. 
Each contribution is weighted by the
corresponding individual average S/N ratio. The co-addition is
a geometric mean to preserve the continuum logarithmic slope
\citep[see][]{Francis1991}. 
Once all spectra are co-added a sliding gaussian filter
($\sigma = 2$ \AA\ ) is applied to the composite to improve the S/N
ratio.  

\begin{figure*}
      \centering
      \resizebox{\linewidth}{!}{\includegraphics{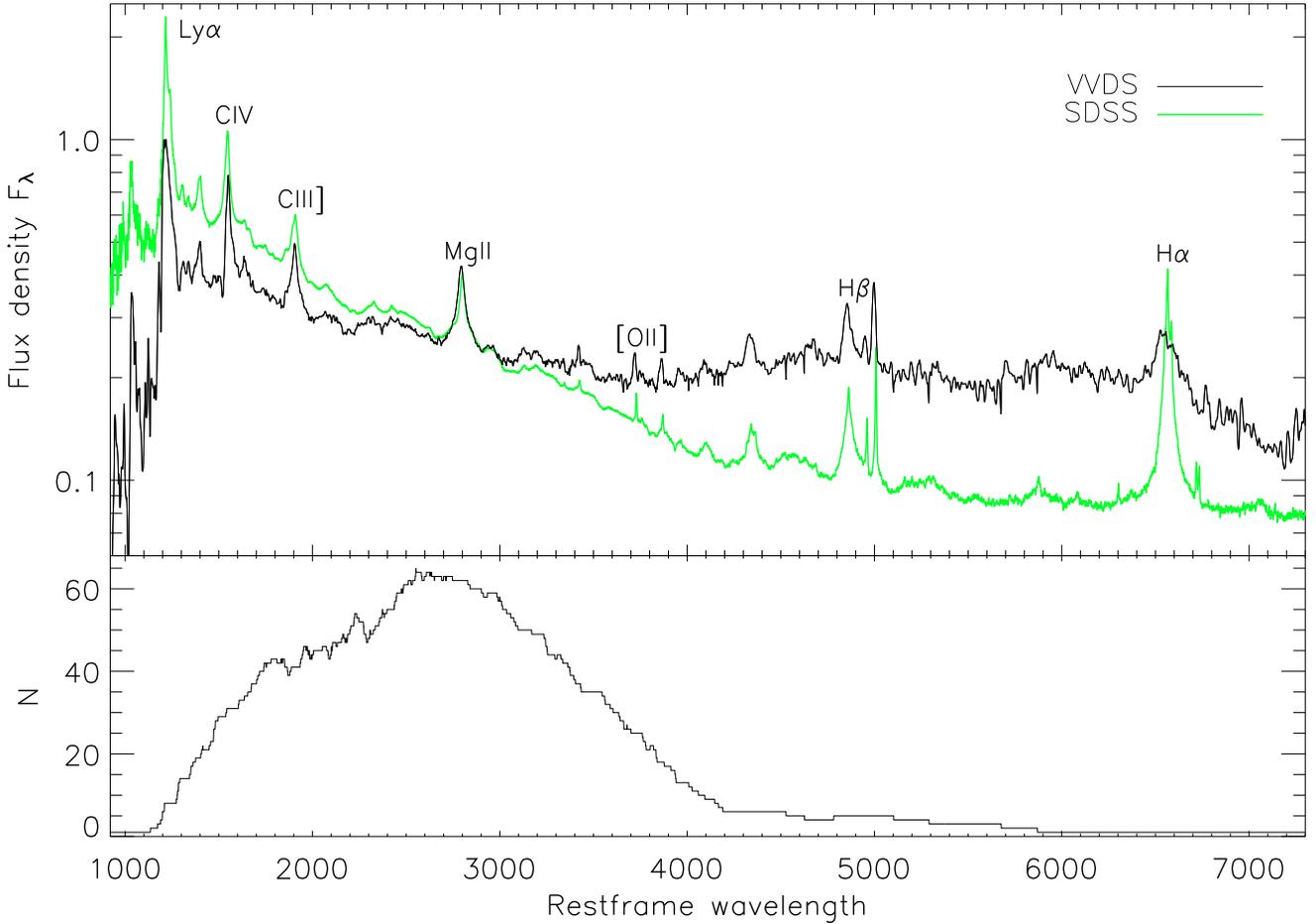}}
        \caption{Composite spectrum . Top panel: Composite spectrum of the
      VVDS BLAGN with secure 
      redshift (black) compared with the SDSS composite spectrum (gray or
      green in electronic version). Bottom panel: Number of contributing
      spectra as a function of wavelength.}
      \label{fig:composite}
    \end{figure*}

Individual redshifts are slightly varied so as to maximize the peaks of 
each main emission line, then they are averaged and corrected
to fix the line peaks close to their laboratory wavelengths.
Then the entire procedure is iterated, the convergence being 
quite fast after 2 or 3 iterations.
All 115 BLAGN spectra with secure redshifts but one have been used to
  generate the composite. In particular, BAL spectra are included.
  The spectrum of object 000029274 at redshift 0.7352, is strongly affected by
  the residual noise of the fringing pattern and contributes to a wavelength
  range covered by only 5 other objects. For this reason it was excluded to
  generate the composite.
The final composite obtained is shown in Figure \ref{fig:composite} along with
the number of contributing spectra.
For comparison the SDSS composite spectrum obtained with the
same method over 2200 spectra is plotted over our composite
\citep{VandenBerk2001}. 
  
 %__________________________________________________ Figure composite
 %__________________________________________________
 
 The maximum number of individual spectra contributing to the
 composite is 64 near 2570 \AA .   
 At both wavelength ends, a few objects only contribute to the
 composite spectrum. 
The signal to noise is $\sim 60$ in the $2000-4000$ \AA\ range and goes down to
 $\sim 30$ at larger wavelength.
We measured the EW of the main emission lines in the composite using the IRAF
 package \footnote{http://iraf.noao.edu/}.
Broad emission lines have been fitted by a Lorentzian profile while narrow emission
 lines are have been by a Gaussian. CIV and Lya lines have been deblended from their
 absorption component.
The resulting EW values are given Table \ref{tab:lines} and are compared to those of
 the SDSS composite spectrum.
 
 Significant differences are seen in the EW of $Ly\alpha$
 and $H\alpha$. 
 The reason for such discrepancies is likely to be essentially due to
 the large variety of EWs for the same
 line observed in QSO spectra, since very few objects are
 contributing at both ends of the VVDS composite spectrum, 5 or less below
 1200 \AA \  and above 5050 \AA. 

We interpret the larger EW we obtain for the CIV line as a consequence of
the Baldwin effect \citep{Baldwin1977}, our objects being a factor up to 100 times fainter
than the ones of the SDSS sample.

The overall continuum shape of our composite spectrum is redder than that
of the SDSS composite and this is particularly true at $\lambda > 3000$ \AA.
This can again be an indication that the host galaxy contamination is
reddening our composite spectrum at long wavelength.
This result is consistent with the discussion made in the section about the
red colors we observe for our fainter objects.

 To further check this interpretation, a new composite was computed excluding
 the extended objects found in our morphological analysis above, for which we
 expect the contamination of the host galaxy to be stronger.
This composite of point-like BLAGN is indeed found to have a bluer
continuum than our composite with the complete sample. 

Finally, we subtracted the continuum of the SDSS composite spectrum
from that of the VVDS composite; emission lines have been previously fitted
and subtracted from the composite spectrum.

\begin{figure}
      \centering
      \resizebox{\linewidth}{!}{\includegraphics{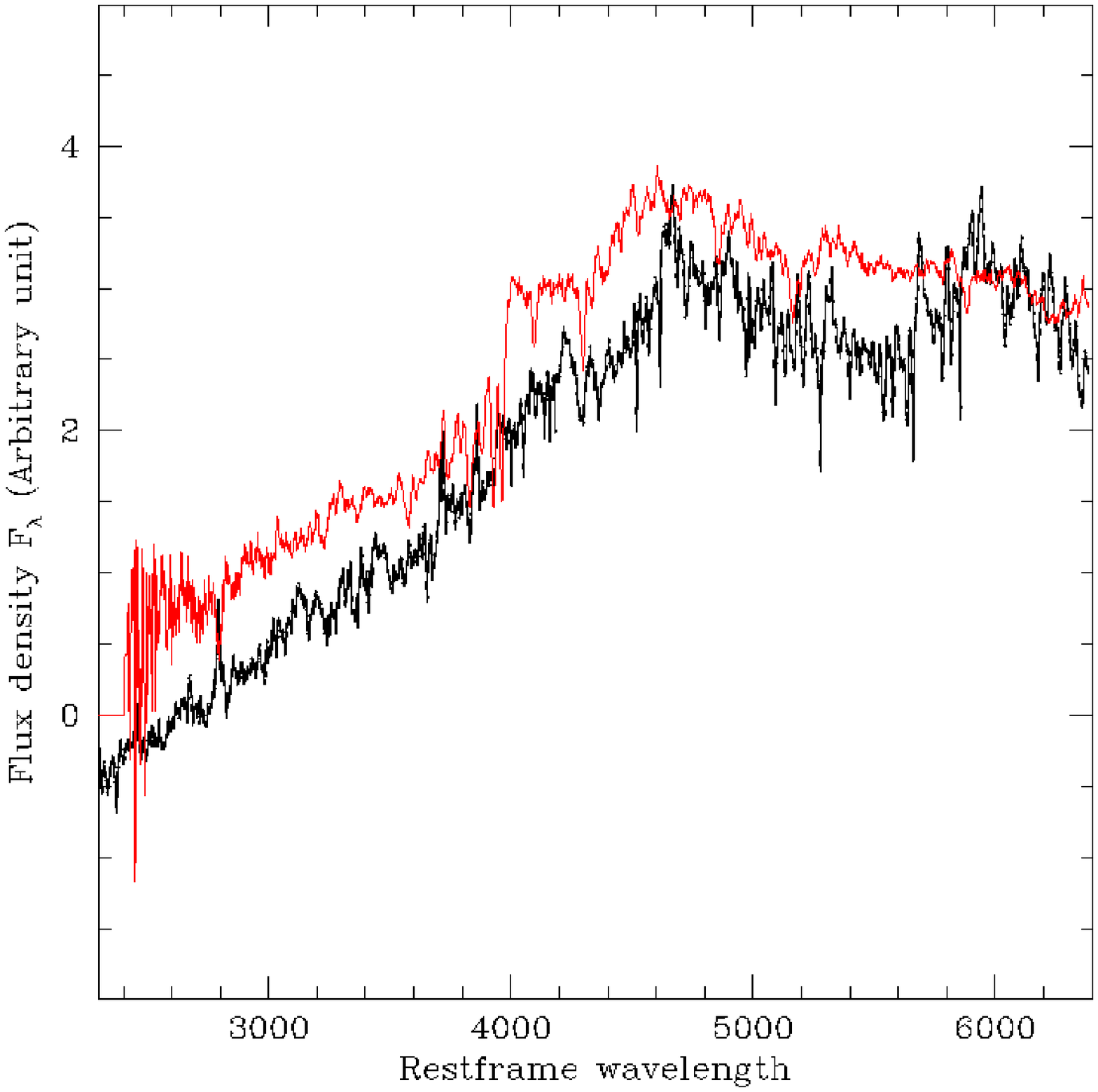}}
        \caption{Difference between VVDS-AGN composite spectrum and the SDSS
      one. For comparison, the red line show the template of early type
        galaxies as derived from the VVDS data} 
      \label{fig:compaHG}
    \end{figure}

The resulting difference is shown in Figure  \ref{fig:compaHG} together with
the spectrum of the template of early type galaxies as derived from the VVDS
data \citep{Contini2006}. The good agreement of the overall shape of the two 
spectra further strengthens our conclusion about the probable significant 
contamination of the red part of our composite spectrum by emission from the
host galaxies.   
 
 %__________________________________________________ Figure comparison with galaxy
 %__________________________________________________
 %__________________________________________________ Table lines
   \begin{table}
      \caption[]{Rest-frame equivalent widths and full width half maximum of
        the VVDS composite spectrum compare to the SDSS composite spectrum.}
         \label{tab:lines}
         $$
         \begin{array}{cccccccc}
            \hline \hline
             \noalign{\smallskip}
line           & \lambda_0 &   EW (VVDS)  & EW (SDSS) \\\hline
Ly _\alpha +N\,V & 1216+1240 &137.4 & 92.91 \\
Si\,IV+O\,IV \textbf{]}   & 1396+1402 &  7.9 &  8.13 \\
C\,IV          & 1549      & 46.0 & 23.78 \\
C\,III \textbf{]}        & 1909      & 25.5 & 21.19 \\
Mg\,II         & 2798      & 40.3 & 32.28 \\
$$\textbf{[}$$ O\,II \textbf{]}   & 3727      &  3.4 &  1.56 \\
$$\textbf{[}$$ Ne\,III \textbf{]} & 3869      &  2.7 &  1.38 \\
H _\delta      & 4103      &  5.0 &  5.05 \\
H _\gamma      & 4342      & 21.8 & 12.62 \\ 
H _\beta       & 4861      & 33.8 & 46.21 \\
$$\textbf{[}$$ O\,III \textbf{]}  & 4959      &  4.9 &  3.50 \\
$$\textbf{[}$$ O\,III \textbf{]}  & 5007      & 16.8 & 13.23 \\
H _\alpha      & 6562      &137.5 &194.52 \\
\hline
           \noalign{\smallskip}
            \hline
        \end{array}
$$
\end{table}
%--------------------------------------------------

\section{Summary and conclusion}
%\subsection{selection function}
This paper describes a complete sample of 130 BLAGN selected from
the VVDS first epoch spectroscopic catalog down to $I_{AB} = 24$. 
It is the first spectroscopic BLAGN catalog of this size at such faint magnitudes,
purely magnitude limited and free of preselection biases.

%\subsection{counts}
The VVDS deep and wide complete samples contain 74 and 56 BLAGN respectively in this first 
release. A total of about 250 to 300 BLAGN are expected when the VVDS will be completed.
We measure cumulative surface densities of $ 472 \pm 48$ BLAGN per sq. deg. with $I_{AB} \le 24$.

%\subsection{z distribution}
The redshift distribution ranges from 0 to 5 while the mean redshift in both the WIDE and DEEP samples is 1.8.
So, by pushing our magnitude limit from $I_{AB} = 22.5$ to
$I_{AB} = 24$ the result is not to increase the mean
redshift of the sample, but rather to explore the faint end of the
luminosity function at all redshifts. 

%\subsection{Morphological and color analysis}
By comparing the $u-g$, $g-r$ color distribution of our AGN population
  with $z < 2.3$ and $I_{AB} < 22.5$ to the complete VVDS photometric catalog
  of stellar-like objects,  we find that $\sim 35\%$ of the AGN present in our
  sample would be missed by the usual UV excess and morphological
  criteria used for the preselection of optical QSO candidates in this
  redshift range.
Most of the extended VVDS BLAGN are below the redshift $z=1.6$,
a redshift range where 42\% of VVDS BLAGN are extended. 

The VVDS BLAGN have redder colors at the faint end of the
luminosity function. Although we cannot exclude that the intrinsic spectrum
of BLAGN is redder at faint luminosity or that the redder color it is due to
the presence of dust, there are evidences that this effect is due to
contamination by the continuum of the host galaxy at faint magnitudes. 
Indeed, 35\% of the VVDS BLAGN have magnitudes fainter than $M_B=-23$.

%\subsection{spectra analysis}
This contamination is also seen in the composite spectrum
obtained by co-adding the individual spectra in their rest frame.
A comparison of the VVDS and SDSS composite QSO spectra
shows that  the VVDS continuum is significantly redder than the SDSS one,
especially at long wavelengths.

%\subsection{prospects}
In the context of the study of the VVDS luminosity function and its evolution 
\citep{Bongiorno2006}, our BLAGN sample has two interesting properties.
First, it is free of biases in the redshift range $2<z<3$, which will
help to shed light on the change of QSO evolution suspected to happen in
this redshift range. 
Second, since no bias against low luminosity AGN is present in this sample,
this will make easier the comparison of the evolution of low and high
luminosity AGN, allowing a better understanding of the QSO-Seyfert
connection. 

\begin{acknowledgements}
This research has been developed within the framework of the VVDS
consortium.
This work has been partially supported by the
CNRS-INSU and its Programme National de Cosmologie (France),
and by Italian Ministry (MIUR) grants
COFIN2000 (MM02037133) and COFIN2003 (2003020150).
The VLT-VIMOS observations have been carried out on guaranteed
time (GTO) allocated by the European Southern Observatory (ESO)
to the VIRMOS consortium, under a contractual agreement between the
Centre National de la Recherche Scientifique of France, heading
a consortium of French and Italian institutes, and ESO,
to design, manufacture and test the VIMOS instrument.
Based on observations obtained with MegaPrime/MegaCam, a joint project of CFHT
and CEA/DAPNIA, at the Canada-France-Hawaii Telescope (CFHT) which is operated
by the National Research Council (NRC) of Canada, the Institut National des
Sciences de l'Univers of the Centre National de la Recherche Scientifique
(CNRS) of France, and the University of Hawaii. 
This work is based in part on data products produced at TERAPIX and
the Canadian Astronomy Data Center as part of the Canada-France-Hawaii
Telescope Legacy Survey, a collaborative project of the National
Research Council (NRC) of Canada and of the Centre National de la
Recherche Scientifique (CNRS) of France. 
\end{acknowledgements}

\bibliographystyle{aa}

\appendix
\section{Catalog Tables}
\label{tab}

%__________________________________________________ Table deep AB
\begin{longtable}{lrrrlccccc}
  \caption[]{\label{tab:AB} BLAGN with secure redshift (flag 14 and 13).}   \\
  \hline\hline
  $ Object \, ID $&$ \hfill  \alpha_{J2000} \hfill $&$
  \hfill  \delta_{J2000} \hfill $&$ \hfill   z \hfill $&$ flag $&$ I_{AB} $&$
  B_{AB} $&$ V_{AB} $&$ R_{AB} $&$Morphology^{(*)}$\\
  \hline
  \endfirsthead
  \caption{continued.}\\
  \hline\hline
  $ Object \, ID $&$ \hfill  \alpha_{J2000} \hfill $&$
  \hfill  \delta_{J2000} \hfill $&$ \hfill   z \hfill $&$ flag $&$ I_{AB} $&$
  B_{AB} $&$ V_{AB} $&$ R_{AB} $&$Morphology^{(*)}$\\
  \hline
  \endhead
  \hline
  \endfoot
   CDFS: & deep \, mode &          10\, AGN \hfill & & & \\ \hline
000037103 & $03^{h} 32^{m} 37.47^{s} $ & $-27^{o} 40^{m} 00.33^{s} $ & $ 0.6656$ & $14$ & $21.84$ & -- $$ & -- $$ & -- $$ & -- $$ \\
000037399 & $03^{h} 32^{m} 38.14^{s} $ & $-27^{o} 39^{m} 45.02^{s} $ & $ 0.8366$ & $14$ & $20.44$ & -- $$ & -- $$ & -- $$ & -- $$ \\
000073509 & $03^{h} 32^{m} 02.47^{s} $ & $-27^{o} 46^{m} 00.53^{s} $ & $ 1.6199$ & $14^{(S)}$ & $23.62$ & -- $$ & -- $$ & -- $$ & -- $$ \\
000023526 & $03^{h} 32^{m} 43.25^{s} $ & $-27^{o} 49^{m} 14.38^{s} $ & $ 1.9199$ & $14^{(S)}$ & $22.17$ & -- $$ & -- $$ & -- $$ & -- $$ \\
000028880 & $03^{h} 33^{m} 03.62^{s} $ & $-27^{o} 45^{m} 18.97^{s} $ & $ 1.2574$ & $14$ & $22.71$ & -- $$ & -- $$ & -- $$ & -- $$ \\
000029274 & $03^{h} 32^{m} 30.23^{s} $ & $-27^{o} 45^{m} 04.75^{s} $ & $ 0.7352$ & $14$ & $21.62$ & -- $$ & -- $$ & -- $$ & -- $$ \\
000018607 & $03^{h} 32^{m} 18.26^{s} $ & $-27^{o} 52^{m} 41.42^{s} $ & $ 2.8010$ & $14$ & $23.91$ & -- $$ & -- $$ & -- $$ & -- $$ \\
000025363 & $03^{h} 32^{m} 59.85^{s} $ & $-27^{o} 47^{m} 48.42^{s} $ & $ 2.5673$ & $14$ & $21.95$ & -- $$ & -- $$ & -- $$ & -- $$ \\
000033629 & $03^{h} 32^{m} 25.17^{s} $ & $-27^{o} 42^{m} 19.05^{s} $ & $ 1.6207$ & $14$ & $22.14$ & -- $$ & -- $$ & -- $$ & -- $$ \\
000031947 & $03^{h} 32^{m} 00.37^{s} $ & $-27^{o} 43^{m} 19.85^{s} $ & $ 1.0401$ & $14^{(S)}$ & $22.12$ & -- $$ & -- $$ & -- $$ & -- $$ \\
\hline
 0226-04 & deep \, mode &          56\, AGN \hfill & & & \\ \hline
020176565 & $02^{h} 25^{m} 28.06^{s} $ & $-04^{o} 36^{m} 41.59^{s} $ & $ 1.5039$ & $14$ & $23.24$ & $23.62$ & $23.61$ & $23.26$ & -- $$ \\
020158952 & $02^{h} 26^{m} 17.81^{s} $ & $-04^{o} 39^{m} 08.50^{s} $ & $ 0.8738$ & $14$ & $21.41$ & $22.82$ & $22.16$ & $21.99$ & extended $$ \\
020086859 & $02^{h} 26^{m} 29.62^{s} $ & $-04^{o} 49^{m} 14.41^{s} $ & $ 1.1921$ & $13$ & $20.97$ & $22.98$ & $21.97$ & $21.43$ & point-like $$ \\
020213000 & $02^{h} 26^{m} 47.88^{s} $ & $-04^{o} 31^{m} 35.20^{s} $ & $ 1.2250$ & $13$ & $21.44$ & $22.30$ & $22.02$ & $21.76$ & point-like $$ \\
020212038 & $02^{h} 26^{m} 08.40^{s} $ & $-04^{o} 31^{m} 43.15^{s} $ & $ 2.2082$ & $14^{(F)}$ & $21.46$ & $22.91$ & $22.27$ & $22.02$ & extended $$ \\
020131908 & $02^{h} 26^{m} 51.04^{s} $ & $-04^{o} 42^{m} 56.55^{s} $ & $ 2.7813$ & $14$ & $22.34$ & $23.08$ & $22.75$ & $22.76$ & extended $$ \\
020210524 & $02^{h} 27^{m} 07.55^{s} $ & $-04^{o} 32^{m} 02.98^{s} $ & $ 1.5150$ & $14$ & $20.41$ & $21.03$ & $20.90$ & $20.64$ & point-like $$ \\
020120394 & $02^{h} 26^{m} 59.92^{s} $ & $-04^{o} 44^{m} 30.32^{s} $ & $ 1.6120$ & $14$ & $20.38$ & $21.27$ & $20.96$ & $20.78$ & point-like $$ \\
020114448 & $02^{h} 27^{m} 00.99^{s} $ & $-04^{o} 45^{m} 16.83^{s} $ & $ 1.6140$ & $13$ & $22.24$ & $23.39$ & $23.35$ & $22.97$ & point-like $$ \\
020118986 & $02^{h} 26^{m} 54.53^{s} $ & $-04^{o} 44^{m} 37.72^{s} $ & $ 3.3018$ & $14$ & $23.57$ & $25.42$ & $23.98$ & $23.97$ & -- $$ \\
020118483 & $02^{h} 27^{m} 36.06^{s} $ & $-04^{o} 44^{m} 41.89^{s} $ & $ 1.2606$ & $13$ & $22.86$ & $23.82$ & $23.44$ & $23.30$ & -- $$ \\
020188089 & $02^{h} 25^{m} 25.68^{s} $ & $-04^{o} 35^{m} 09.45^{s} $ & $ 2.1384$ & $14$ & $20.66$ & $21.12$ & $20.94$ & $20.95$ & point-like $$ \\
020147295 & $02^{h} 25^{m} 29.19^{s} $ & $-04^{o} 40^{m} 44.16^{s} $ & $ 1.5562$ & $14^{(F)}$ & $22.59$ & $23.87$ & $23.44$ & $22.84$ & -- $$ \\
020169816 & $02^{h} 25^{m} 45.04^{s} $ & $-04^{o} 37^{m} 35.95^{s} $ & $ 3.5893$ & $14^{(F)}$ & $22.14$ & $24.67$ & $22.95$ & $22.64$ & point-like $$ \\
020190479 & $02^{h} 25^{m} 45.55^{s} $ & $-04^{o} 34^{m} 45.18^{s} $ & $ 0.1524$ & $14$ & $21.33$ & $21.99$ & $21.55$ & $21.64$ & point-like $$ \\
020268754 & $02^{h} 26^{m} 09.63^{s} $ & $-04^{o} 24^{m} 37.74^{s} $ & $ 2.7187$ & $14$ & $20.59$ & $20.56$ & $20.60$ & $20.75$ & point-like $$ \\
020164607 & $02^{h} 25^{m} 32.46^{s} $ & $-04^{o} 38^{m} 18.63^{s} $ & $ 2.9220$ & $14$ & $23.11$ & $23.92$ & $23.46$ & $23.21$ & -- $$ \\
020179116 & $02^{h} 25^{m} 34.98^{s} $ & $-04^{o} 36^{m} 16.46^{s} $ & $ 3.3080$ & $14^{(F)}$ & $23.89$ & $25.18$ & $23.94$ & $23.84$ & -- $$ \\
020237445 & $02^{h} 25^{m} 57.38^{s} $ & $-04^{o} 28^{m} 46.04^{s} $ & $ 1.2138$ & $14$ & $22.43$ & $23.92$ & $23.53$ & $23.06$ & extended $$ \\
020223153 & $02^{h} 26^{m} 17.52^{s} $ & $-04^{o} 30^{m} 29.27^{s} $ & $ 1.4777$ & $14^{(F)}$ & $20.65$ & $21.18$ & $21.05$ & $20.94$ & point-like $$ \\
020163018 & $02^{h} 26^{m} 45.20^{s} $ & $-04^{o} 38^{m} 30.58^{s} $ & $ 1.3208$ & $14^{(F)}$ & $23.11$ & $23.89$ & $23.62$ & $23.29$ & -- $$ \\
020180665 & $02^{h} 26^{m} 45.46^{s} $ & $-04^{o} 36^{m} 15.43^{s} $ & $ 3.2619$ & $14$ & $18.14$ & $21.09$ & $19.32$ & $19.03$ & point-like $$ \\
020177875 & $02^{h} 26^{m} 53.87^{s} $ & $-04^{o} 36^{m} 27.21^{s} $ & $ 1.6821$ & $13$ & $22.53$ & $24.01$ & $23.65$ & $23.36$ & -- $$ \\
020234610 & $02^{h} 26^{m} 58.99^{s} $ & $-04^{o} 29^{m} 06.02^{s} $ & $ 2.1645$ & $13$ & $23.87$ & $26.05$ & $25.11$ & $24.74$ & -- $$ \\
020159510 & $02^{h} 27^{m} 09.85^{s} $ & $-04^{o} 39^{m} 02.21^{s} $ & $ 1.9309$ & $14$ & $21.97$ & $22.73$ & $22.41$ & $22.44$ & point-like $$ \\
020218399 & $02^{h} 27^{m} 31.34^{s} $ & $-04^{o} 30^{m} 50.26^{s} $ & $ 2.2255$ & $14$ & $22.19$ & $22.91$ & $22.39$ & $22.53$ & point-like $$ \\
020254511 & $02^{h} 27^{m} 36.93^{s} $ & $-04^{o} 26^{m} 31.30^{s} $ & $ 1.7466$ & $14$ & $20.66$ & $20.76$ & $20.84$ & $20.85$ & point-like $$ \\
020243922 & $02^{h} 27^{m} 47.33^{s} $ & $-04^{o} 27^{m} 53.20^{s} $ & $ 1.1203$ & $14$ & $21.29$ & $22.02$ & $21.71$ & $21.60$ & point-like $$ \\
020165108 & $02^{h} 26^{m} 59.85^{s} $ & $-04^{o} 38^{m} 12.68^{s} $ & $ 1.3219$ & $13$ & $23.09$ & $24.70$ & $24.04$ & $23.78$ & -- $$ \\
020179225 & $02^{h} 27^{m} 02.15^{s} $ & $-04^{o} 36^{m} 15.96^{s} $ & $ 1.3860$ & $13$ & $22.39$ & $23.90$ & $23.50$ & $23.01$ & point-like $$ \\
020195823 & $02^{h} 27^{m} 24.10^{s} $ & $-04^{o} 33^{m} 55.72^{s} $ & $ 2.4250$ & $14^{(F)}$ & $23.38$ & $24.19$ & $23.61$ & $23.85$ & -- $$ \\
020254576 & $02^{h} 25^{m} 27.23^{s} $ & $-04^{o} 26^{m} 31.02^{s} $ & $ 3.8527$ & $14$ & $21.15$ & $23.33$ & $21.78$ & $21.44$ & point-like $$ \\
020200020 & $02^{h} 25^{m} 50.40^{s} $ & $-04^{o} 33^{m} 24.00^{s} $ & $ 2.7373$ & $13$ & $21.91$ & $21.95$ & $22.10$ & $21.73$ & point-like $$ \\
020329650 & $02^{h} 26^{m} 08.71^{s} $ & $-04^{o} 16^{m} 34.53^{s} $ & $ 1.0498$ & $14$ & $20.84$ & $21.52$ & $21.14$ & $20.96$ & point-like $$ \\
020232397 & $02^{h} 26^{m} 26.04^{s} $ & $-04^{o} 29^{m} 27.88^{s} $ & $ 1.6280$ & $14^{(F)}$ & $22.69$ & $23.72$ & $23.31$ & $23.15$ & -- $$ \\
020461765 & $02^{h} 26^{m} 35.95^{s} $ & $-04^{o} 23^{m} 21.81^{s} $ & $ 3.2831$ & $14$ & $22.89$ & $24.62$ & $23.01$ & $22.87$ & -- $$ \\
020466135 & $02^{h} 26^{m} 46.99^{s} $ & $-04^{o} 18^{m} 37.56^{s} $ & $ 1.5806$ & $14$ & $21.13$ & $21.29$ & $21.56$ & $21.05$ & point-like $$ \\
020467962 & $02^{h} 26^{m} 59.17^{s} $ & $-04^{o} 16^{m} 55.89^{s} $ & $ 3.3247$ & $13$ & $23.59$ & $25.22$ & $23.96$ & $23.80$ & -- $$ \\
020461459 & $02^{h} 27^{m} 04.25^{s} $ & $-04^{o} 23^{m} 37.77^{s} $ & $ 1.8211$ & $13$ & $21.60$ & $23.50$ & $22.92$ & $22.41$ & point-like $$ \\
020465339 & $02^{h} 27^{m} 06.44^{s} $ & $-04^{o} 19^{m} 24.30^{s} $ & $ 3.2852$ & $14$ & $21.09$ & $21.77$ & $21.31$ & $20.99$ & point-like $$ \\
020467628 & $02^{h} 27^{m} 04.06^{s} $ & $-04^{o} 17^{m} 09.77^{s} $ & $ 1.3582$ & $13$ & $21.35$ & $21.97$ & $21.82$ & $21.61$ & point-like $$ \\
020205812 & $02^{h} 27^{m} 23.84^{s} $ & $-04^{o} 32^{m} 31.69^{s} $ & $ 2.8922$ & $14$ & $23.87$ & $24.77$ & $24.36$ & $24.52$ & -- $$ \\
020208084 & $02^{h} 27^{m} 29.24^{s} $ & $-04^{o} 32^{m} 27.51^{s} $ & $ 2.2850$ & $14^{(F)}$ & $19.04$ & $19.34$ & $19.10$ & $19.12$ & point-like $$ \\
020277536 & $02^{h} 27^{m} 53.85^{s} $ & $-04^{o} 23^{m} 20.10^{s} $ & $ 3.6260$ & $14$ & $23.55$ & $25.25$ & $23.88$ & $23.60$ & -- $$ \\
020278210 & $02^{h} 27^{m} 40.00^{s} $ & $-04^{o} 23^{m} 17.43^{s} $ & $ 1.7574$ & $13$ & $21.36$ & $22.68$ & $21.87$ & $22.28$ & point-like $$ \\
020239945 & $02^{h} 27^{m} 31.14^{s} $ & $-04^{o} 28^{m} 22.83^{s} $ & $ 2.4247$ & $14^{(F)}$ & $22.90$ & $23.65$ & $23.13$ & $23.14$ & -- $$ \\
020351846 & $02^{h} 26^{m} 30.84^{s} $ & $-04^{o} 13^{m} 26.09^{s} $ & $ 3.5680$ & $14^{(F)}$ & $22.80$ & $25.05$ & $23.71$ & $23.24$ & -- $$ \\
020367106 & $02^{h} 26^{m} 34.71^{s} $ & $-04^{o} 11^{m} 33.98^{s} $ & $ 1.3973$ & $14^{(F)}$ & $22.42$ & $23.28$ & $23.02$ & $22.55$ & extended $$ \\
020351277 & $02^{h} 25^{m} 57.41^{s} $ & $-04^{o} 13^{m} 39.43^{s} $ & $ 0.6061$ & $14$ & $19.83$ & $20.60$ & $20.39$ & $20.32$ & extended $$ \\
020258622 & $02^{h} 26^{m} 20.06^{s} $ & $-04^{o} 25^{m} 54.51^{s} $ & $ 1.3386$ & $14$ & $22.74$ & $24.45$ & $23.84$ & $23.46$ & -- $$ \\
020286836 & $02^{h} 26^{m} 22.17^{s} $ & $-04^{o} 22^{m} 21.62^{s} $ & $ 2.0060$ & $14$ & $18.51$ & $19.28$ & $19.11$ & $19.05$ & point-like $$ \\
020291309 & $02^{h} 26^{m} 31.23^{s} $ & $-04^{o} 21^{m} 28.87^{s} $ & $ 1.9930$ & $14$ & $22.89$ & $23.95$ & $23.56$ & $23.81$ & -- $$ \\
020465540 & $02^{h} 26^{m} 44.48^{s} $ & $-04^{o} 19^{m} 16.76^{s} $ & $ 2.7372$ & $14$ & $23.58$ & $24.15$ & $23.94$ & $23.81$ & -- $$ \\
020302785 & $02^{h} 26^{m} 24.63^{s} $ & $-04^{o} 20^{m} 02.14^{s} $ & $ 2.2357$ & $14$ & $21.01$ & $21.42$ & $21.26$ & $21.34$ & point-like $$ \\
020364478 & $02^{h} 26^{m} 49.41^{s} $ & $-04^{o} 11^{m} 53.30^{s} $ & $ 1.1573$ & $14$ & $21.74$ & $22.94$ & $22.74$ & $22.25$ & extended $$ \\
020463196 & $02^{h} 27^{m} 00.65^{s} $ & $-04^{o} 21^{m} 49.00^{s} $ & $ 1.3875$ & $14$ & $23.27$ & $24.49$ & $24.42$ & $23.91$ & -- $$ \\
\hline
 1003+01 & wide \, mode &          18 \, AGN \hfill & & & \\ \hline
100122852 & $10^{h} 02^{m} 11.17^{s} $ & $+01^{o} 22^{m} 28.58^{s} $ & $ 1.8007$ & $14$ & $19.86$ & -- $$ & -- $$ & -- $$ & point-like $$ \\
100110223 & $10^{h} 02^{m} 48.14^{s} $ & $+01^{o} 20^{m} 02.29^{s} $ & $ 1.8255$ & $13$ & $21.30$ & -- $$ & -- $$ & -- $$ & point-like $$ \\
100210521 & $10^{h} 03^{m} 27.33^{s} $ & $+01^{o} 35^{m} 50.91^{s} $ & $ 1.1723$ & $14$ & $21.59$ & $22.05$ & $21.73$ & $21.56$ & point-like $$ \\
100139500 & $10^{h} 02^{m} 57.37^{s} $ & $+01^{o} 25^{m} 40.38^{s} $ & $ 1.2478$ & $13$ & $20.96$ & -- $$ & -- $$ & -- $$ & point-like $$ \\
100126868 & $10^{h} 03^{m} 08.80^{s} $ & $+01^{o} 23^{m} 16.56^{s} $ & $ 2.3302$ & $14$ & $20.34$ & -- $$ & -- $$ & -- $$ & point-like $$ \\
100105943 & $10^{h} 03^{m} 46.33^{s} $ & $+01^{o} 19^{m} 11.04^{s} $ & $ 3.5553$ & $13$ & $21.15$ & -- $$ & -- $$ & -- $$ & point-like $$ \\
100290682 & $10^{h} 03^{m} 11.33^{s} $ & $+01^{o} 47^{m} 01.56^{s} $ & $ 1.5487$ & $14^{(F)}$ & $21.45$ & -- $$ & -- $$ & -- $$ & extended $$ \\
100327652 & $10^{h} 03^{m} 13.81^{s} $ & $+01^{o} 52^{m} 13.97^{s} $ & $ 1.2173$ & $14$ & $22.39$ & $23.85$ & $23.43$ & $22.90$ & extended $$ \\
100232259 & $10^{h} 03^{m} 30.37^{s} $ & $+01^{o} 38^{m} 51.18^{s} $ & $ 1.7647$ & $14$ & $21.09$ & $21.37$ & $21.30$ & $21.33$ & point-like $$ \\
100190464 & $10^{h} 04^{m} 25.14^{s} $ & $+01^{o} 33^{m} 07.74^{s} $ & $ 1.0760$ & $14$ & $21.66$ & $22.98$ & $22.38$ & $22.41$ & extended $$ \\
100168207 & $10^{h} 04^{m} 36.55^{s} $ & $+01^{o} 30^{m} 05.86^{s} $ & $ 2.7152$ & $14$ & $22.40$ & $22.38$ & $21.84$ & $22.14$ & point-like $$ \\
100113463 & $10^{h} 04^{m} 07.25^{s} $ & $+01^{o} 20^{m} 38.90^{s} $ & $ 1.8436$ & $14$ & $20.47$ & -- $$ & -- $$ & -- $$ & point-like $$ \\
100123590 & $10^{h} 04^{m} 46.72^{s} $ & $+01^{o} 22^{m} 39.10^{s} $ & $ 2.0963$ & $13$ & $21.47$ & -- $$ & -- $$ & -- $$ & point-like $$ \\
100343840 & $10^{h} 04^{m} 32.08^{s} $ & $+01^{o} 54^{m} 24.12^{s} $ & $ 2.3666$ & $14^{(F)}$ & $19.75$ & $20.68$ & $20.40$ & $20.15$ & point-like $$ \\
100338914 & $10^{h} 04^{m} 13.45^{s} $ & $+01^{o} 53^{m} 41.38^{s} $ & $ 1.1584$ & $14$ & $19.73$ & $20.15$ & $19.80$ & $19.78$ & extended $$ \\
100245809 & $10^{h} 04^{m} 00.36^{s} $ & $+01^{o} 40^{m} 45.74^{s} $ & $ 3.0789$ & $14$ & $21.62$ & $22.71$ & $22.01$ & $21.94$ & point-like $$ \\
100241696 & $10^{h} 04^{m} 11.84^{s} $ & $+01^{o} 40^{m} 06.47^{s} $ & $ 1.1112$ & $13$ & $21.95$ & $22.59$ & $21.89$ & $22.19$ & point-like $$ \\
100451895 & $10^{h} 04^{m} 38.01^{s} $ & $+02^{o} 09^{m} 25.07^{s} $ & $ 1.7806$ & $13$ & $20.03$ & $20.46$ & $20.31$ & $20.36$ & point-like $$ \\
\hline
 2217+00 & wide \, mode &           31\, AGN \hfill  & & & \\ \hline
220586430 & $22^{h} 14^{m} 34.82^{s} $ & $+00^{o} 19^{m} 24.18^{s} $ & $ 1.0285$ & $14$ & $20.35$ & -- $$ & -- $$ & -- $$ & point-like $$ \\
220568559 & $22^{h} 14^{m} 43.23^{s} $ & $+00^{o} 14^{m} 16.29^{s} $ & $ 1.4980$ & $13$ & $22.10$ & -- $$ & -- $$ & -- $$ & point-like $$ \\
220566905 & $22^{h} 14^{m} 02.39^{s} $ & $+00^{o} 13^{m} 49.58^{s} $ & $ 1.5285$ & $13$ & $22.28$ & -- $$ & -- $$ & -- $$ & point-like $$ \\
220554336 & $22^{h} 14^{m} 44.17^{s} $ & $+00^{o} 10^{m} 02.54^{s} $ & $ 0.4470$ & $14$ & $21.02$ & -- $$ & -- $$ & -- $$ & point-like $$ \\
220001963 & $22^{h} 13^{m} 51.58^{s} $ & $+00^{o} 25^{m} 01.30^{s} $ & $ 2.6801$ & $14$ & $22.19$ & -- $$ & -- $$ & -- $$ & point-like $$ \\
220010371 & $22^{h} 14^{m} 28.40^{s} $ & $+00^{o} 27^{m} 32.40^{s} $ & $ 3.6952$ & $14$ & $21.79$ & -- $$ & -- $$ & -- $$ & point-like $$ \\
220056847 & $22^{h} 14^{m} 48.77^{s} $ & $+00^{o} 41^{m} 16.67^{s} $ & $ 3.0015$ & $14$ & $21.38$ & -- $$ & -- $$ & -- $$ & point-like $$ \\
220567825 & $22^{h} 15^{m} 08.48^{s} $ & $+00^{o} 14^{m} 04.38^{s} $ & $ 1.1601$ & $13$ & $21.07$ & -- $$ & -- $$ & -- $$ & point-like $$ \\
220576817 & $22^{h} 15^{m} 09.17^{s} $ & $+00^{o} 16^{m} 42.38^{s} $ & $ 3.0957$ & $14$ & $21.81$ & -- $$ & -- $$ & -- $$ & point-like $$ \\
220536609 & $22^{h} 15^{m} 31.65^{s} $ & $+00^{o} 04^{m} 18.31^{s} $ & $ 0.4970$ & $14$ & $21.10$ & -- $$ & -- $$ & -- $$ & extended $$ \\
220041929 & $22^{h} 15^{m} 09.54^{s} $ & $+00^{o} 36^{m} 39.11^{s} $ & $ 1.4751$ & $13$ & $18.24$ & -- $$ & -- $$ & -- $$ & extended $$ \\
220090821 & $22^{h} 15^{m} 46.25^{s} $ & $+00^{o} 50^{m} 58.51^{s} $ & $ 1.8326$ & $13$ & $20.40$ & -- $$ & -- $$ & -- $$ & point-like $$ \\
220082140 & $22^{h} 15^{m} 32.40^{s} $ & $+00^{o} 48^{m} 36.29^{s} $ & $ 1.8484$ & $14^{(F)}$ & $20.58$ & -- $$ & -- $$ & -- $$ & point-like $$ \\
220055529 & $22^{h} 15^{m} 54.10^{s} $ & $+00^{o} 40^{m} 55.47^{s} $ & $ 3.5941$ & $13$ & $21.42$ & -- $$ & -- $$ & -- $$ & point-like $$ \\
220567863 & $22^{h} 16^{m} 27.06^{s} $ & $+00^{o} 14^{m} 02.32^{s} $ & $ 2.1610$ & $14$ & $20.80$ & -- $$ & -- $$ & -- $$ & point-like $$ \\
220567224 & $22^{h} 16^{m} 44.02^{s} $ & $+00^{o} 13^{m} 48.54^{s} $ & $ 5.0042$ & $14$ & $20.07$ & -- $$ & $22.90$ & -- $$ & point-like $$ \\
220591287 & $22^{h} 16^{m} 49.05^{s} $ & $+00^{o} 20^{m} 46.27^{s} $ & $ 1.2968$ & $14^{(F)}$ & $22.30$ & $23.47$ & $23.32$ & -- $$ & extended $$ \\
220580912 & $22^{h} 15^{m} 56.66^{s} $ & $+00^{o} 17^{m} 52.28^{s} $ & $ 3.0432$ & $13$ & $22.11$ & -- $$ & -- $$ & -- $$ & point-like $$ \\
220107230 & $22^{h} 16^{m} 56.10^{s} $ & $+00^{o} 56^{m} 00.77^{s} $ & $ 1.0937$ & $13$ & $21.64$ & -- $$ & -- $$ & -- $$ & point-like $$ \\
220575888 & $22^{h} 17^{m} 36.55^{s} $ & $+00^{o} 16^{m} 23.09^{s} $ & $ 3.0755$ & $14$ & $21.25$ & $22.16$ & $21.50$ & -- $$ & point-like $$ \\
220556037 & $22^{h} 17^{m} 05.53^{s} $ & $+00^{o} 10^{m} 19.85^{s} $ & $ 2.7422$ & $14$ & $19.46$ & $20.52$ & $20.29$ & -- $$ & point-like $$ \\
220542377 & $22^{h} 17^{m} 10.42^{s} $ & $+00^{o} 06^{m} 04.72^{s} $ & $ 1.3097$ & $13$ & $21.12$ & $22.24$ & $21.95$ & -- $$ & point-like $$ \\
220554600 & $22^{h} 17^{m} 36.64^{s} $ & $+00^{o} 10^{m} 05.86^{s} $ & $ 1.3689$ & $14^{(F)}$ & $20.65$ & $21.21$ & $21.00$ & -- $$ & point-like $$ \\
220544855 & $22^{h} 17^{m} 39.71^{s} $ & $+00^{o} 06^{m} 52.80^{s} $ & $ 2.2934$ & $14^{(F)}$ & $20.91$ & $21.97$ & $21.61$ & -- $$ & extended $$ \\
220044408 & $22^{h} 17^{m} 34.47^{s} $ & $+00^{o} 37^{m} 33.52^{s} $ & $ 2.9096$ & $14^{(F)}$ & $21.18$ & $21.65$ & $21.54$ & -- $$ & point-like $$ \\
220093875 & $22^{h} 17^{m} 48.64^{s} $ & $+00^{o} 51^{m} 50.39^{s} $ & $ 1.3365$ & $14^{(F)}$ & $21.83$ & -- $$ & -- $$ & -- $$ & point-like $$ \\
220081925 & $22^{h} 18^{m} 00.42^{s} $ & $+00^{o} 48^{m} 31.41^{s} $ & $ 1.2167$ & $13$ & $21.68$ & $22.79$ & $22.40$ & -- $$ & extended $$ \\
220609820 & $22^{h} 18^{m} 29.04^{s} $ & $+00^{o} 20^{m} 24.32^{s} $ & $ 1.4794$ & $14^{(F)}$ & $21.52$ & $22.04$ & $21.86$ & -- $$ & extended $$ \\
220610034 & $22^{h} 18^{m} 14.20^{s} $ & $+00^{o} 20^{m} 49.73^{s} $ & $ 1.5135$ & $14^{(F)}$ & $20.54$ & $20.77$ & $20.86$ & -- $$ & extended $$ \\
220613346 & $22^{h} 18^{m} 33.73^{s} $ & $+00^{o} 27^{m} 09.76^{s} $ & $ 1.2530$ & $14^{(F)}$ & $20.39$ & $21.63$ & $21.25$ & -- $$ & extended $$ \\
220098629 & $22^{h} 18^{m} 01.51^{s} $ & $+00^{o} 53^{m} 19.83^{s} $ & $ 2.5790$ & $14^{(F)}$ & $21.71$ & -- $$ & -- $$ & -- $$ & point-like $$ \\

%%% Local Variables: 
%%% mode: latex
%%% TeX-master: t
%%% End: 

\end{longtable}
\begin{description}
  \item[$^{(*)}$] Morphological classification for objects with $I_{AB} <
    22.5$. Objects in the VVDS-CDFS field are not classified. See Section \ref{sec:morph};
  \item[$^{(F)}$] AGN re-observed in the 3800-6500 range with FORS1 in our follow-up program;
  \item[$^{(S)}$] Redshift confirmed by matching the catalog from \cite{Szokoly2004}.
\end{description}
%--------------------------------------------------
%__________________________________________________ Table C
\begin{table}
\caption[]{\label{tab:C} AGN with a single emission line detected (flag 19).}
\rotatebox{90}{
\begin{tabular}{lrrrlccccc}
  \hline\hline
  $Object \, ID $&$   \alpha_{J2000}  $&
  $  \delta_{J2000}  $&$  \lambda_{BL} 
  $&$ z \, solutions $&$ I_{AB} $&$
  B_{AB} $&$ V_{AB} $&$ R_{AB} $&$Morphology^{(*)}$\\
  \hline
   CDFS: & deep\, mode  \hfill&           2\, AGN \hfill &&&&&&&\\ \hline
000031270 &$ 03^{h} 32^{m} 57.74^{s}  $&$ -27^{o} 43^{m} 50.12^{s}  $&$ 7319. $&$ \,  0.1150^a\,  1.6150^b   $&$ 23.52$ & -- $$ & -- $$ & -- $$ & -- $$ \\
000017025 &$ 03^{h} 31^{m} 54.30^{s}  $&$ -27^{o} 53^{m} 49.58^{s}  $&$ 6500. $&$  \,  1.3220^b\,  2.4050^c\,  3.1960^d $&$ 23.70$ & -- $$ & -- $$ & -- $$ & -- $$ \\
\hline 0226-04 : & deep\, mode  \hfill&           6\, AGN \hfill &&&&&&&\\ \hline
020137737 &$ 02^{h} 26^{m} 47.76^{s}  $&$ -04^{o} 42^{m} 04.06^{s}  $&$ 6320. $&$  \,  1.2580^b\,  2.3110^c\,  3.0800^d $&$ 23.78$ & $24.58$ & $24.14$ & $24.16$ & -- $$ \\
020225567 &$ 02^{h} 27^{m} 06.42^{s}  $&$ -04^{o} 30^{m} 14.34^{s}  $&$ 6558. $&$  \,  1.3430^b\,  2.4350^c  $&$ 23.07$ & $24.65$ & $24.26$ & $23.74$ & -- $$ \\
020281035 &$ 02^{h} 26^{m} 12.30^{s}  $&$ -04^{o} 22^{m} 51.63^{s}  $&$ 6805. $&$ \,  0.0370^a\,  1.4310^b\,  2.5650^c\,  3.3930^d $&$ 23.93$ & $25.06$ & $24.39$ & $24.07$ & -- $$ \\
020375508 &$ 02^{h} 25^{m} 48.99^{s}  $&$ -04^{o} 10^{m} 28.04^{s}  $&$ 5963. $&$  \,  1.1300^b\,  2.1240^c\,  2.8500^d $&$ 22.54$ & $23.25$ & $22.98$ & $23.04$ & -- $$ \\
020293248 &$ 02^{h} 26^{m} 25.92^{s}  $&$ -04^{o} 21^{m} 12.73^{s}  $&$ 7335. $&$ \,  0.1180^a\,  1.6210^b   $&$ 23.25$ & $24.79$ & $24.54$ & $24.13$ & -- $$ \\
020469530 &$ 02^{h} 26^{m} 49.92^{s}  $&$ -04^{o} 15^{m} 17.44^{s}  $&$ 6701. $&$ \,  0.0210^a\,  1.3940^b\,  2.5100^c\,  3.3260^d $&$ 23.50$ & $24.69$ & $24.35$ & $24.28$ & -- $$ \\
\hline  2217+00: & wide\, mode  \hfill&           7\, AGN \hfill &&&&&&&\\ \hline
220593613 &$ 22^{h} 14^{m} 11.61^{s}  $&$ +00^{o} 21^{m} 29.15^{s}  $&$ 6893. $&$ \,  0.0500^a\,  1.4630^b\,  2.6110^c  $&$ 21.90$ & -- $$ & -- $$ & -- $$ & extended $$ \\
220056092 &$ 22^{h} 13^{m} 53.81^{s}  $&$ +00^{o} 41^{m} 06.90^{s}  $&$ 6100. $&$  \,  1.1790^b\,  2.1950^c  $&$ 22.08$ & -- $$ & -- $$ & -- $$ & point-like $$ \\
220583713 &$ 22^{h} 15^{m} 34.70^{s}  $&$ +00^{o} 18^{m} 42.01^{s}  $&$ 7352. $&$ \,  0.1200^a\,  1.6270^b   $&$ 21.88$ & -- $$ & -- $$ & -- $$ & point-like $$ \\
220548678 &$ 22^{h} 15^{m} 02.71^{s}  $&$ +00^{o} 08^{m} 10.56^{s}  $&$ 6766. $&$ \,  0.0310^a\,  1.4170^b\,  2.5440^c  $&$ 21.97$ & -- $$ & -- $$ & -- $$ & point-like $$ \\
220023681 &$ 22^{h} 17^{m} 46.44^{s}  $&$ +00^{o} 31^{m} 26.58^{s}  $&$ 6956. $&$ \,  0.0600^a\,  1.4850^b\,  2.6440^c  $&$ 21.71$ & $24.33$ & $23.25$ & -- $$ & point-like $$ \\
220551735 &$ 22^{h} 18^{m} 05.78^{s}  $&$ +00^{o} 09^{m} 12.66^{s}  $&$ 8056. $&$ \,  0.2270^a\,  1.8780^b   $&$ 21.12$ & $21.73$ & $21.72$ & -- $$ & point-like $$ \\
220234909 &$ 22^{h} 18^{m} 13.40^{s}  $&$ +00^{o} 48^{m} 54.05^{s}  $&$ 6407. $&$  \,  1.2890^b\,  2.3560^c\,  3.1360^d $&$ 21.88$ & $23.40$ & $22.55$ & -- $$ & extended $$ \\

  \hline
  \multicolumn{10}{l}{\textbf{$^{(*)}$} Morphological classification for
  objects with $I_{AB} < 22.5$. Objects in the VVDS-CDFS field are not
  classified. See Section \ref{sec:morph}}\\
  \multicolumn{10}{l}{\textbf{$^a$} Emission line identified as H$\alpha$.}\\
  \multicolumn{10}{l}{\textbf{$^b$} Emission line identified as MgII.}     \\
  \multicolumn{10}{l}{\textbf{$^c$} Emission line identified as C\,III].}     \\
  \multicolumn{10}{l}{\textbf{$^d$} Emission line identified as CIV .}     \\
\end{tabular}}
\end{table}
%--------------------------------------------------
\end{document}